\documentclass{aastex631}
\usepackage{cleveref} 
\usepackage{subfigure}
\usepackage{appendix}
\usepackage{graphicx}

\begin{document}

\title{A Neural Network–Based Tool for Rapid Orbital Solution of Contact Binary Systems}

\correspondingauthor{Kai Li}
\email{kaili@sdu.edu.cn}

\author{Li-Heng Wang}
\affiliation{Shandong Key Laboratory of Optical Astronomy and Solar-Terrestrial Environment, School of Space Science and Technology, Institute of Space Sciences, Shandong University, Weihai, Shandong 264209, China}

\author{Kai Li}
\affiliation{Shandong Key Laboratory of Optical Astronomy and Solar-Terrestrial Environment, School of Space Science and Technology, Institute of Space Sciences, Shandong University, Weihai, Shandong 264209, China}

\author{Ya-Ni Guo}
\affiliation{Shandong Key Laboratory of Optical Astronomy and Solar-Terrestrial Environment, School of Space Science and Technology, Institute of Space Sciences, Shandong University, Weihai, Shandong 264209, China}



\begin{abstract}
With the rapid growth of large-scale time-domain surveys, vast amounts of photometric and spectroscopic data are continuously being released. A substantial number of contact binaries have been identified, accompanied by both light curves and radial velocity observations. Traditional modeling of these datasets relies on computationally intensive physical models, which limits their applicability to large-scale analyses. To this end, this paper proposes the use of a fully connected neural network and a SIREN  \citep{siren} model to rapidly generate radial velocity curves. After training, the models can rapidly generate radial velocity curves with high accuracy, significantly reducing computational costs and enabling efficient large-scale studies. This method was further integrated with the light curve computation framework developed by \cite{TCW2025},  allowing simultaneous fitting of radial velocities and light curves. To assess the reliability of the proposed model, synthetic observational datasets were constructed and used for parameter estimation. The deviations between the derived parameters and the ground-truth values were analyzed quantitatively, enabling a comprehensive evaluation of the model’s accuracy.

\end{abstract}



\section{Introduction} \label{sec:intro}

Binary systems are widespread in the Universe and are commonly categorized, based on their Roche geometry, into detached, semi-detached, and contact configurations \citep{Kopal1955}. Contact binaries are systems in which both components fill or overfill their Roche lobes and share a common envelope. Many such systems, particularly W UMa-type binaries, consist of late-type stars. For these systems, despite substantial differences in mass between the two components, their effective temperatures are typically very similar \citep{Kuiper1941}.  These distinctive properties, together with the strong and continuous interactions between the two components, make contact binaries valuable laboratories for investigating stellar structure, mass transfer, and binary evolution. Accurate determinations of their orbital and physical parameters can provide important observational constraints on models of stellar structure and binary evolution, and further contribute to studies of stellar population synthesis and galaxy evolution.

Despite substantial advances in recent years, during which a considerable number of contact binaries have been modeled to determine their physical properties, important questions remain unresolved. In particular, comprehensive and statistically robust samples of accurately determined physical parameters are needed to constrain phenomena such as the mass-ratio limit \citep{Rasio1995, Lifang2006, LiK2024}, the short-period cutoff \citep{Rucinski1992, Jiang2012, Qi-Qi2019}, and spot evolution \citep{mengzibin, wangsirui}. 

In recent years, the rapid expansion of sky survey facilities and continuous advances in instrumentation have led to an exponential growth in astronomical data volumes. In response, a variety of efficient data-driven methods have been developed, including deep learning techniques for variable star classification \citep{Classification2022}, automated galaxy morphology classification \citep{Zhirui2022}, and neural network–based inference of physical parameters for contact binaries \citep{TCW2025, TCW2, dingxu}. These approaches have significantly improved the efficiency of large-scale data analysis and facilitated investigations into the evolutionary processes of stellar systems and galaxies. For contact binaries, light and radial velocity curves are traditionally modeled using the Wilson–Devinney \cite[W–D;][]{Wilson1971, Wilson1979, Wilson1990} code or PHysics Of Eclipsing BinariEs \cite[PHOEBE;][]{pras2016, pras2005}. Detailed modeling of close binary systems, including proximity effects such as tidal distortion, gravity darkening, and irradiation, is typically carried out using well-established Roche-geometry-based codes. Several modeling codes are currently available for this purpose, among which W-D and PHOEBE are the most widely used \citep{2021ApJSwdphoebe}. 
However, performing binary star modeling with W-D or PHOEBE is computationally time-consuming, particularly in scenarios involving large volumes of observational data. A single model evaluation can often take several minutes, while applying Markov Chain Monte Carlo \cite[MCMC;][]{mcmc}  methods may require several days. 
To address this limitation, we propose two different neural network–based models that can directly and rapidly generate theoretical radial velocity curves from input physical parameters. We further integrate these models with the light-curve inversion framework developed by \cite{TCW2025}, thereby establishing a unified scientific tool for the simultaneous solution of light curves and radial velocity curves.

\section{ Neural Network Model Construction} \label{sec:dam}

\subsection{Dataset Construction }\label{dc}

We first used PHOEBE 2.4 to generate radial velocity curves with \texttt{rv\_method = flux-weighted}, by considering proximity effect, which served as the training dataset.\textbf{ }In order to cover the parameters that may influence the shape of the curves as comprehensively as possible, we selected seven key parameters: the mass ratio ($q$, from 0.05 to 1),  the effective temperature of the primary component ($T_1$, from 4000K to 10000K), the temperature ratio ($T_2$/$T_1$, from 0.7 to 1.2), the period ($P$, from 0.15d to 1d), the semi-major axis ($a$, from 1$R_\odot$ to 10$R_\odot$), the orbital inclination ($i$, from 30$^\circ$ to 90$^\circ$ ) ,  the fill-out factor ($f$, from 0 to 1). Instead of sampling the component radii directly, we sampled the fill-out factor, which determines the degree of contact and therefore constrains the stellar radii through the Roche geometry.
To better unify the scale of the input parameters, we applied the following normalization scheme: sma/10, $i$/90, and $T_1$/7000. 
The gravity-darkening coefficient and bolometric albedo were assigned according to the envelope structure of the stellar components. The physical origin of this treatment lies in the different energy transport mechanisms in convective and radiative envelopes. For stars with convective envelopes, energy transport is dominated by convection, which reduces the dependence of the local surface flux on the effective gravity. Therefore, a lower gravity-darkening coefficient of g=0.32 is adopted following \cite{Lucy1967}. In addition, efficient energy redistribution in convective atmospheres leads to a bolometric albedo of A=0.5 following \cite{Ruci1969}.
For stars with radiative envelopes, the emergent flux is governed by radiative energy transport and follows the \cite{v1924}, resulting in a gravity-darkening coefficient of g=1. The bolometric albedo is correspondingly set to A=1, assuming complete re-radiation of the absorbed energy.
Following previous studies  \cite[e.g., ][]{7200_1, 7200_2, 7200_3}, we use an effective temperature of approximately 7200 K as the empirical boundary between convective and radiative envelopes. If the temperature is below 7200 K, the gravity-darkening and bolometric albedo coefficients are set to g=0.32 and A=0.5, respectively. When the temperature exceeds 7200 K, both parameters are set to 1.
And the atmospheric model proposed by \cite{CF}  was used. Within these ranges, we randomly sampled 50,000 sets of parameters. 

Using PHOEBE, we generated 100 data points at equal intervals within the phase range of 0–1, and produced the corresponding radial velocity curves to form the dataset. 
To evaluate whether 100 uniformly sampled phase points are sufficient to capture the detailed structures of radial velocity curves, especially the localized variations caused by the Rossiter–McLaughlin effect during eclipses, we performed additional sampling tests. Following the radial velocity anomalies induced by the Rossiter–McLaughlin effect reported by \cite{2026Barone}, we specifically examined the phase interval from -0.05 to 0.05, where rapid RV variations may occur. In this interval, we compared the curves sampled with 10 and 100 phase points. The 10-point sampling corresponds to the resolution adopted by our model within this phase range. The 10-point curve was interpolated to 100 points using cubic spline interpolation and compared with the original 100-point curve. We calculated the maximum absolute difference between the interpolated and original data, which was 0.132 $km/s$. Furthermore, we generated RV curves using PHOEBE with 100 and 1000 uniformly distributed phase points. The 100-point curves were interpolated to 1000 points using cubic spline interpolation and compared with the original high-resolution curves. The maximum absolute difference between the interpolated and original curves was 0.0751 $km/s$. For comparison, \cite{wang2019} reported RV precisions of 1.36, 1.08, and 0.91 $km/s$ for LAMOST medium-resolution spectra at S/N ratios of 10, 20, and 50, respectively. Meanwhile, \cite{Rucinski2002} reported an RV precision of approximately 1 $km/s$ for sharp-lined stars observed with their telescope, with generally lower precision for broad-lined binary components. Therefore, the interpolation errors are substantially smaller than the typical  observational RV uncertainties, indicating that the adopted phase resolution can accurately reproduce the local RV variations and that 100 phase points are sufficient to preserve the detailed structures of the simulated RV curves.

During the generation of the synthetic dataset, we noted that some PHOEBE-generated radial velocity curves contained NaN values. These NaNs typically occurred at phases where the spectral information of the eclipsed component was unavailable, and therefore no observational radial velocity measurement could be obtained at these phases. Consequently, predicting a continuous RV value at these locations does not affect practical data analysis, since these phases do not provide valid observational constraints. Since standard fully connected neural network (FCNN) cannot directly handle NaN values during supervised training, we removed the NaN data points from the curves while retaining the remaining valid data points for training. This prevents invalid values from entering the loss calculation. During inference, the neural network predicts the underlying continuous radial velocity function from the input physical parameters and therefore does not generate missing values, even for parameter combinations where PHOEBE may return NaN values at specific phases. The predicted values at these locations represent model-estimated RV values rather than missing observational measurements. Since these points do not correspond to actual observations, they do not affect the results.

\subsection{FCNN And SIREN }\label{fc}

\begin{figure}
    \centering
    \includegraphics[width=1\linewidth]{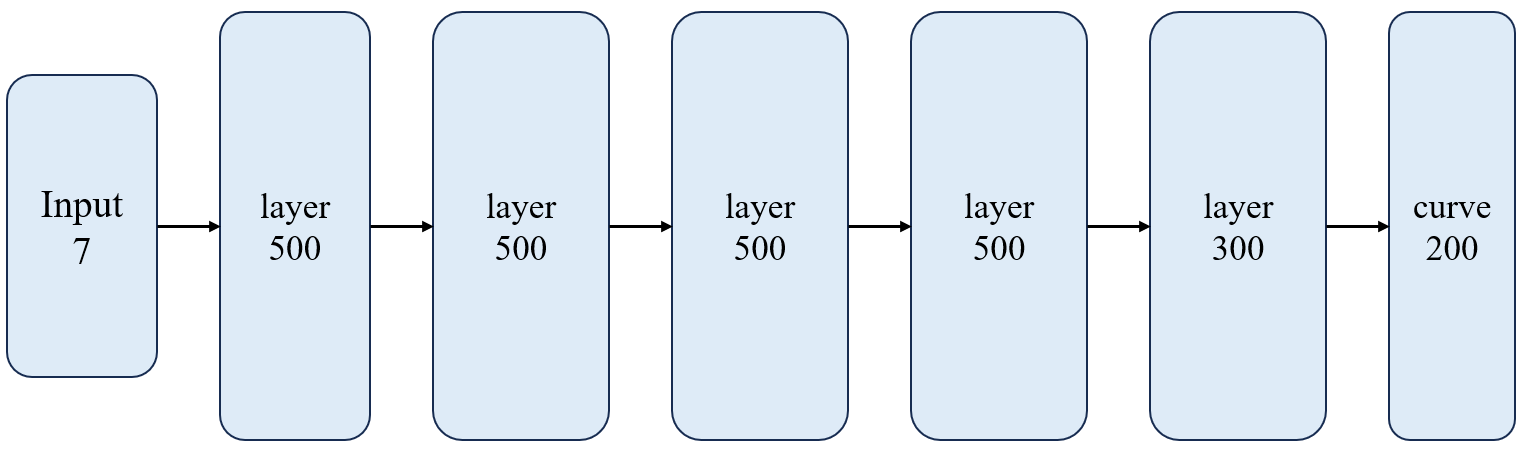}
    \caption{Structure of the FCNN.}
    \label{fig:NNS}
\end{figure}

\begin{figure}
    \centering
    \includegraphics[width=1\linewidth]{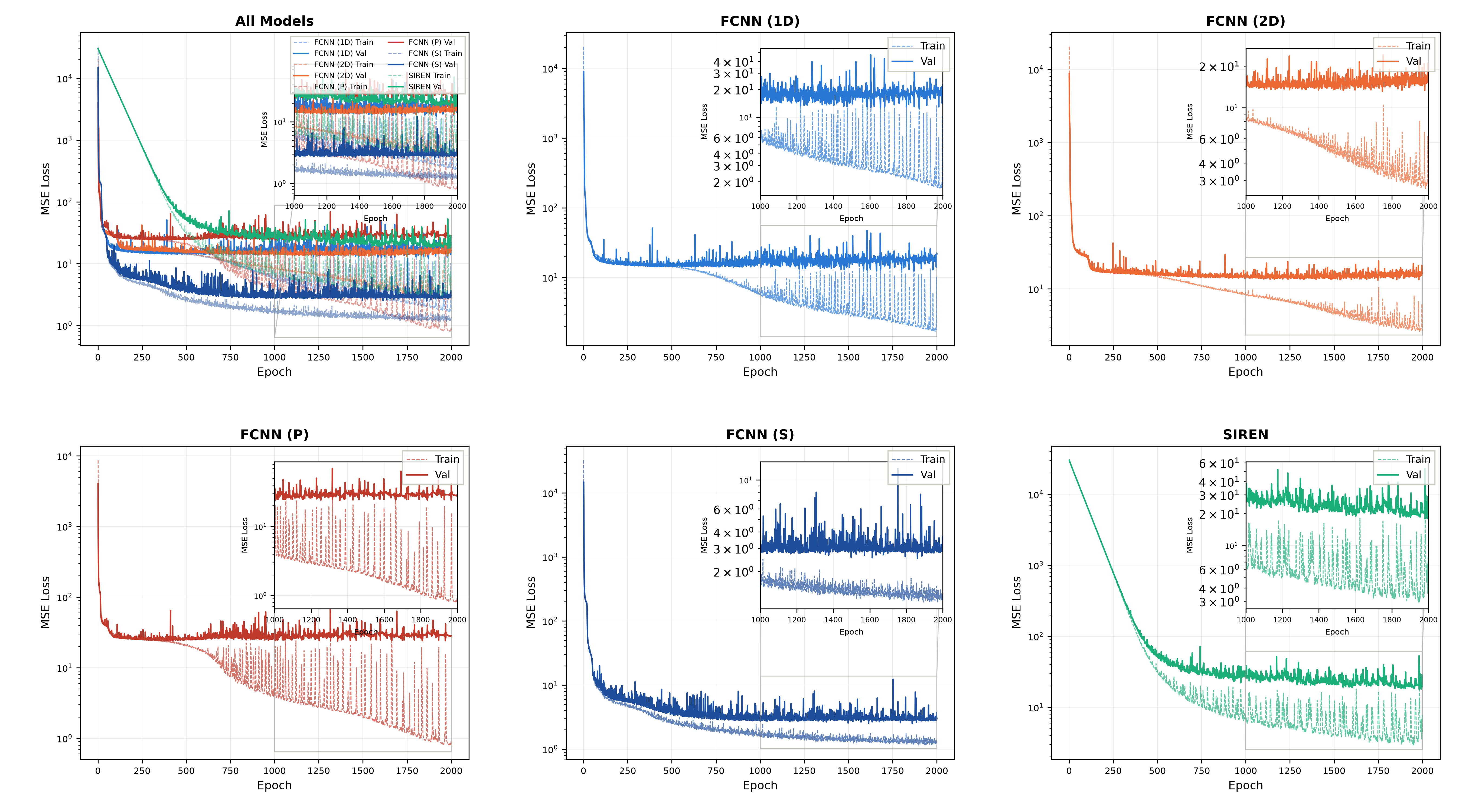}
    \caption{The training and validation loss curves throughout the training process.}
    \label{fig:loss}
\end{figure}

FCNN are widely used for nonlinear function approximation and can learn complex mappings between input and output variables. 
We constructed a FCNN model, whose architecture is illustrated in  \Cref{fig:NNS} . 
Meanwhile, to allow the model to simultaneously predict the radial velocity curves of the primary and secondary components, we explored three alternative output representations. The first representation concatenates the two RV curves into a one-dimensional vector (1D, 1 × 200), where the primary-star RV curve is scaled to the range of 0–1 and the secondary-star RV curve is shifted to the range of 1–2. The second representation preserves the two curves as separate channels by normalizing both RV curves to the range of 0–1 and arranging them into a two-dimensional output (2D, 2 × 100). The third representation consists of two independent FCNNs, one for predicting the radial velocity curve of the primary component and the other for predicting that of the secondary component. Both models use the same input and hidden-layer architecture as the 1D model, with separate output layers designed for the respective components.
The dataset was divided into training and validation sets with a ratio of 8:2, and the batch size for training was set to 64. GPU acceleration was employed to expedite the training process. 
The training process adopted the mean squared error (MSE) as the objective function. The ReLU activation function \citep{relu2015} was applied to the hidden layers, while the output layer employed a linear activation function to allow predictions over the full range of radial velocity values.
Optimization was performed using the Adam optimizer \citep{Adam2014}, which enables adaptive learning rate updates and promotes stable convergence. The initial learning rate was initialized to 0.0001, and the network was trained for a total of 2000 epochs. The training and validation loss curves during the training process are shown in  \Cref{fig:loss}. 

\begin{figure}
    \centering
    \includegraphics[width=1\linewidth]{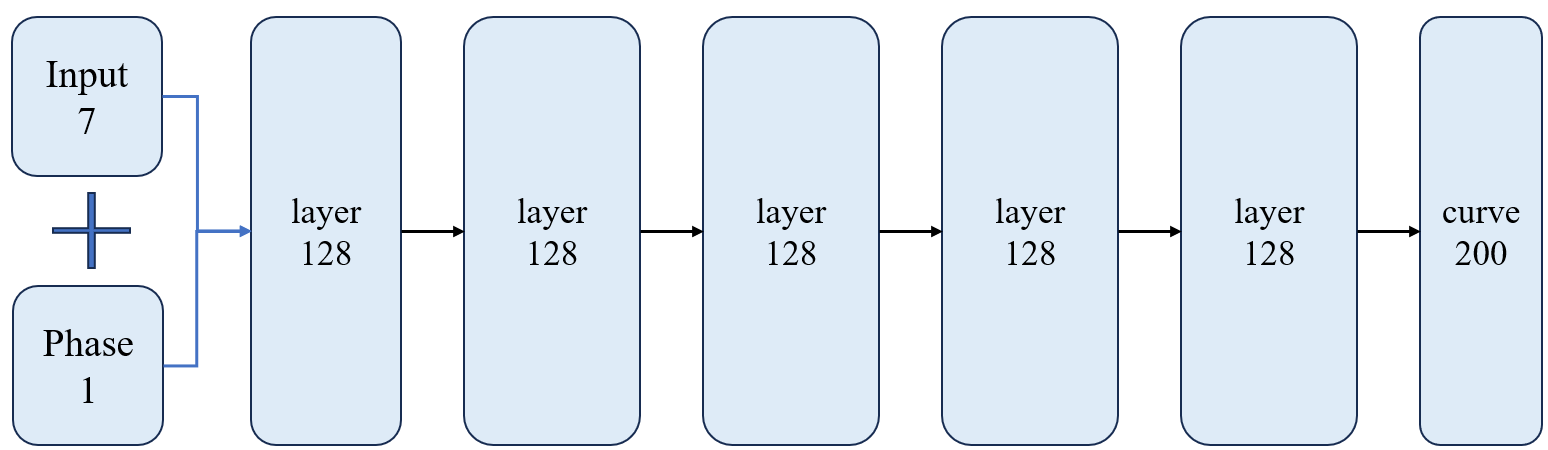}
    \caption{The structure of the SIREN model.}
    \label{fig:siren}
\end{figure}

Although the FCNN adopted in this work provides sufficient representational capacity, exploring more compact architectures is still of practical interest, particularly for applications involving large-scale population studies or repeated inference on large datasets. A more compact model can reduce computational and memory requirements while maintaining comparable predictive performance. Considering that radial velocity curves exhibit strong periodicity and their profiles are approximately sinusoidal, we additionally adopt the Sinusoidal Representation Network \cite[SIREN;][]{siren} architecture. 
SIREN uses sinusoidal activation functions and is particularly well suited for representing continuous periodic functions, while also offering a substantially smaller model size than a FCNN. 
In our model, we set the sine frequency scaling factor to 10. The model introduces the phase coordinate as an additional input variable. For each phase point, the network predicts the corresponding radial velocity value. The model architecture is shown in   \Cref{fig:siren} . 
Our SIREN model is capable of generating the complete radial velocity curve through a single forward pass. Specifically, the 200 phase coordinates are concatenated with the seven physical parameters as input, and all phase points are evaluated simultaneously through batched computation without iterative evaluation over individual phase points. 
Compared with a FCNN, the total number of nodes is greatly reduced. Similarly, the dataset was divided into training and testing sets with a ratio of 8:2, the batch size for training was set to 64, and the network was trained for a total of 2000 epochs.
The training and validation loss curves during the training process are shown in  \Cref{fig:loss}.  Although the training loss continued to decrease during the later stages of training, the validation loss of each model gradually approached a stable level, indicating that further training did not lead to a noticeable improvement in generalization performance but mainly resulted in increased overfitting to the training data. To reduce the impact of this late-stage overfitting on the final model, we saved the model weights corresponding to the minimum validation loss during training and used this model for the subsequent analyses, rather than adopting the weights from the final training epoch.

\subsection{Training Results }\label{fc}

The final model results indicate that the FCNN model is 11 times larger than the SIREN model.
To quantitatively assess model performance, 10,000 synthetic radial velocity curves were generated with PHOEBE across the specified parameter range. These parameter sets were then supplied to the trained neural network to produce predicted curves. The results of the different models are presented in \Cref{tab:tabel_modle}. 

\renewcommand\arraystretch{1.2}
\begin{table*}[t]
\centering
\caption{\centering This table presents the performance of different models on 10,000 randomly generated curves. }
\label{tab:tabel_modle}
\begin{tabular}{lcccccc}
\hline
model& $median\_R^2$ of P& $median\_R^2$ of S& $median\_R^2$ of P+S& $median\_RSS$ of P& $median\_RSS$ of S&
$median\_RSS$ of P+S\\
\hline
1D& 0.999938& 0.999975& 0.999966& 32.9& 60.4& 107.7\\
2D& 0.999917& 0.999969& 0.999957& 47.0& 73.0& 141.0\\
SIREN& 0.999805& 0.999905& 0.999889& 102.9& 208.4& 335.6\\
FCNN(P)& 0.999941& -& -& 31.7& -& -\\
FCNN(S)& -& 0.999970& -& -& 72.1& -\\

FCNN(P+S)& -& -& 0.999961& -& -&120.7\\
\hline
\end{tabular}
\end{table*}

\begin{figure}[htbp]
    \centering
    \centering
  \includegraphics[width=0.7\linewidth]{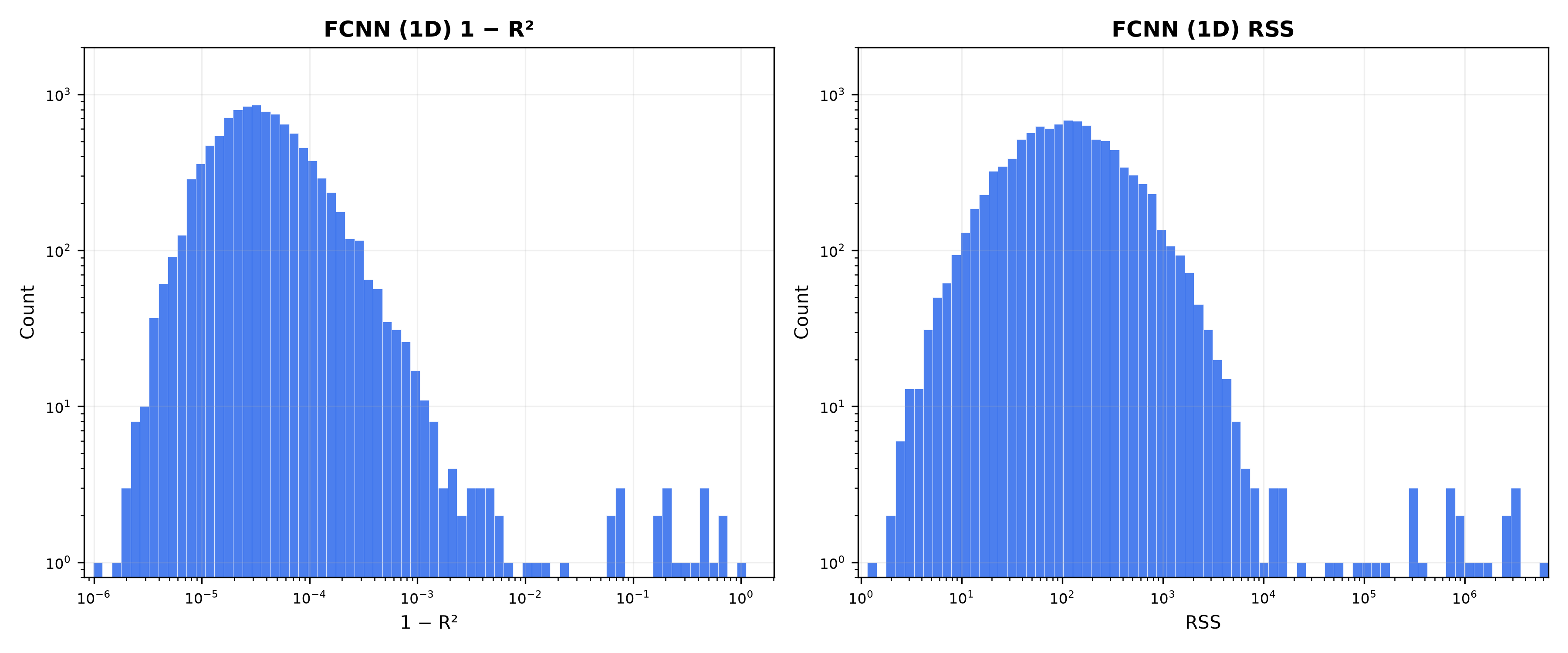}
  \par\medskip
  \includegraphics[width=0.7\linewidth]{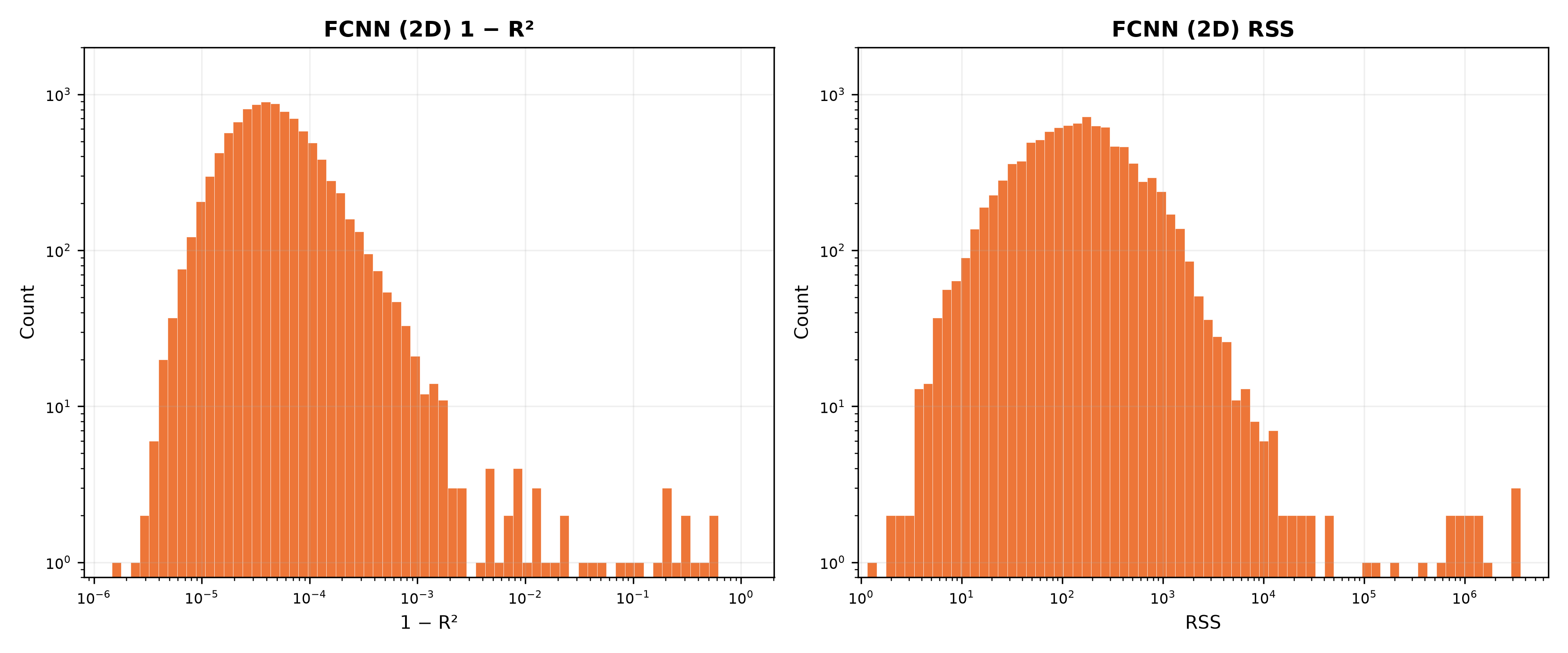}
  \par\medskip
  \includegraphics[width=0.7\linewidth]{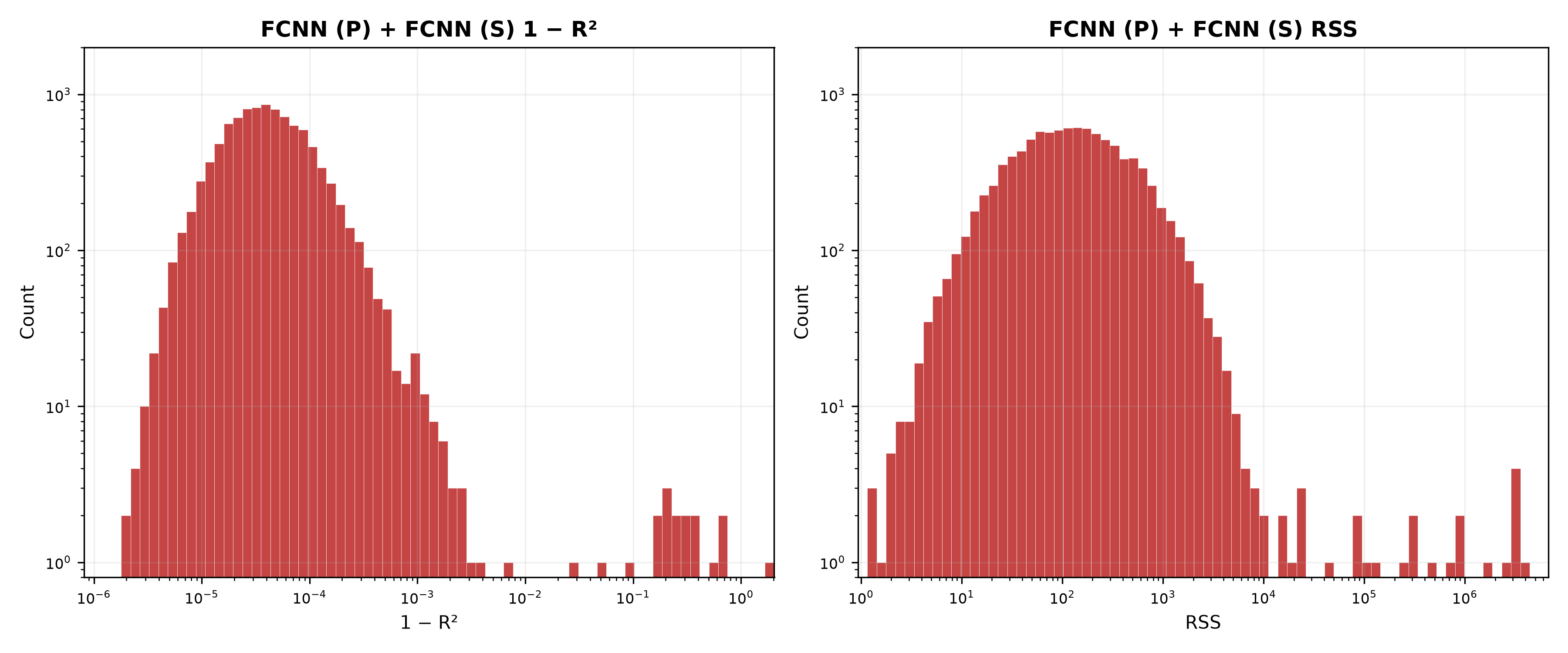}
  \par\medskip
  \includegraphics[width=0.7\linewidth]{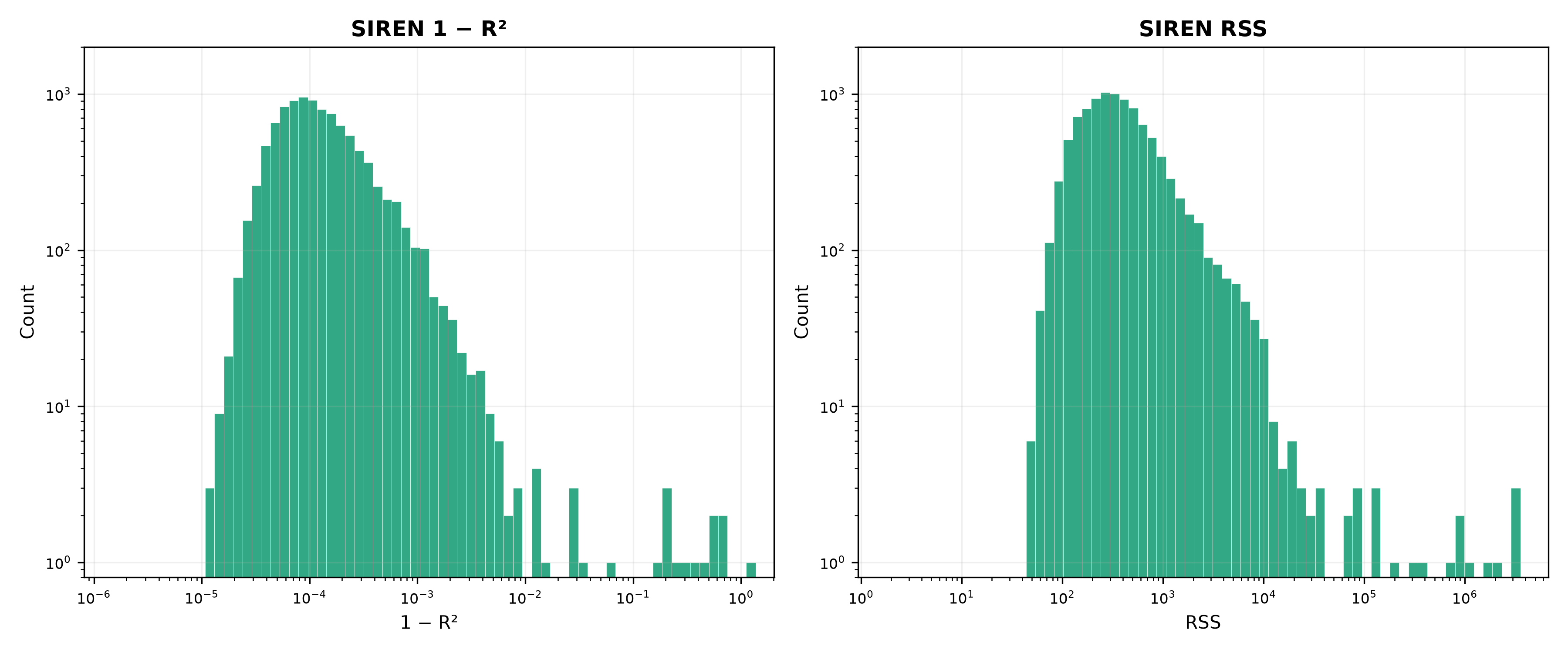}
  \par
    \caption{The figure shows the distributions of $1-R^2$ and residual sum of squares (RSS) for the 1D FCNN, 2D-output FCNN, the combined model consisting of two independent models for the primary and secondary radial velocities, and SIREN models. Both the x- and y-axes are displayed on logarithmic scales with base 10.}
    \label{fig:r2}
\end{figure}

The results show that the performance of the 2D architecture is almost identical to that of the 1D concatenated architecture. We argue that, in a FCNN, individual output neurons are independently connected to the shared hidden representation and do not possess spatial neighboring dependencies. Therefore, the ordering of the output vector does not affect the model performance. We also trained two independent FCNNs, one for the primary component and the other for the secondary component, using the same architecture as the 1D model. Even when each independent model retained the same hidden-layer capacity as the original 1D model, the separate-model approach did not provide a clear overall performance advantage. We suggest that this may be because, although the two curves may benefit from different feature representations, they share the same physical input parameters. Therefore, in the single model, the hidden layers can learn representations that are useful for both components, while the output layer provides separate connections for the two curves. Based on these results, we adopted the simpler 1D architecture for the subsequent analyses.
The distributions of $1-R^2$ and RSS for the predictions of the five models on the 10,000 test curves are presented in \Cref{fig:r2}, with both axes shown on logarithmic scales to better display the overall distributions and their tails. 

We evaluated the computation time of the different models under single-sample inference ($\mathrm{batch\_size}=1$). For generating one radial-velocity curve containing a total of 200 points for the two components, the 1D model required approximately 0.084 ms, while the SIREN model required approximately 0.432 ms. In comparison, a single PHOEBE computation required approximately 3s under the same test conditions. Thus, the SIREN model was approximately 5.2 times slower than the 1D model, whereas PHOEBE required approximately $3.6\times10^4$ times more computation time than the 1D model.
The higher inference cost of the SIREN model compared with the FCNN can be mainly attributed to three factors. First, SIREN takes phase as an input and therefore requires network computations for each phase point, whereas the FCNN can directly generate the complete radial velocity curve through a single forward pass. Second, SIREN employs the sine function as its activation function, which requires additional trigonometric computations compared with the commonly used ReLU activation function, thereby increasing the computational cost. Finally, matrix operations and ReLU used in FCNNs are common operations with relatively mature low-level implementations and optimization strategies, whereas sine operations in SIREN have more limited optimization opportunities, resulting in lower overall execution efficiency. Overall, the SIREN model provides a trade-off between model compactness and computational efficiency, allowing users to select the model according to their specific application requirements.

\section{ Orbital Solution Using the Model} \label{sec:dam2}

We developed a program for simultaneous modeling of radial velocity and light curves of contact binaries by integrating the model described in Section \ref{sec:dam} with the light curve framework of \cite{TCW2025}. The program implements an internal differential evolution \cite[DE;][]{DE}  algorithm and MCMC, enabling efficient batch orbital solutions across large datasets.
For the joint parameter estimation, the light-curve and radial-velocity datasets are treated as conditionally independent given the model parameters. Following the likelihood prescription adopted in PHOEBE 2.4, each measurement is evaluated using its corresponding observational uncertainty. Their likelihoods are then combined with the prior to construct the joint posterior distribution, which is sampled using MCMC.

To evaluate the performance and accuracy of the proposed program, we randomly sampled 10,000 parameter sets. The parameter ranges for the radial velocity were defined as in Section \ref{sec:dam}, while those for the light curves were adopted from \cite{TCW2025}. 
The corresponding light curves and radial velocity curves were then generated using PHOEBE. For batch analysis, the pre-analysis configuration assigned each parameter its maximum allowable range to ensure comprehensive exploration of the parameter space.
 To improve the estimation of the systemic radial velocity ($v_\gamma$), we computed the mean value of the radial velocity curve ($v_{\gamma, mean}$) and defined the search interval for $v_\gamma$ as [$v_{\gamma, mean}$-$100$, $v_{\gamma, mean}+100$]. Since PHOEBE generates theoretical light curves and radial velocity curves, we assigned an uncertainty to each data point to simulate observational errors. The adopted uncertainties covered typical observational ranges, namely [0.0001,0.1] for the light curves and [0.1,50] for the radial velocity curves.
 We first employed DE for parameter exploration. Specifically, we adopted an improved version of the SHADE algorithm, in which four control parameters, including the population size, number of generations, historical memory size ($H$), and greediness parameter ($p$), were explicitly configured. During the batch optimization process, the population size was set to 64, the maximum number of generations was set to 1000, the number of historical memory slots was set to 100, and the elite selection ratio was set to 0.1. In the first optimization stage, the parameter boundaries were set to be consistent with those defined in Section \ref{sec:dam}.
The best solution obtained from the DE optimization was subsequently adopted as the initial state for the second-stage MCMC sampling. In this stage, the number of walkers ($n_{\mathrm{walkers}}$) was set to 32, and the maximum number of iterations was set to 10,000. The autocorrelation time was checked every 100 iterations, and the program was considered to have converged and terminated when $50\tau_{\max}<N$, where $\tau_{\max}$ is the largest autocorrelation time among all parameters and $N$ is the current number of iterations.

\begin{figure*}[htbp]

\subfigure{
\includegraphics[width=9cm,height = 5cm]{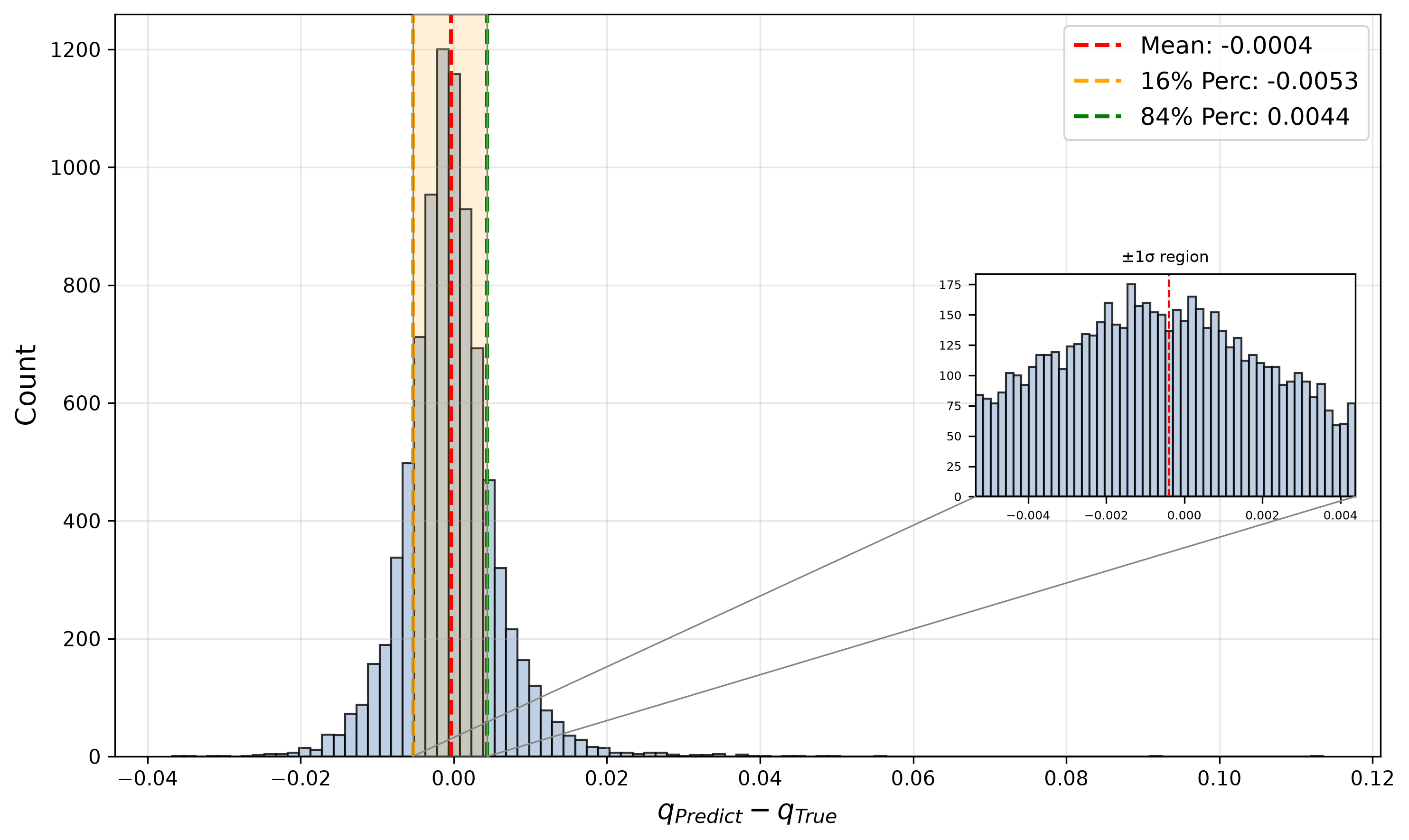}}
\subfigure{
\includegraphics[width=9cm,height = 5cm]{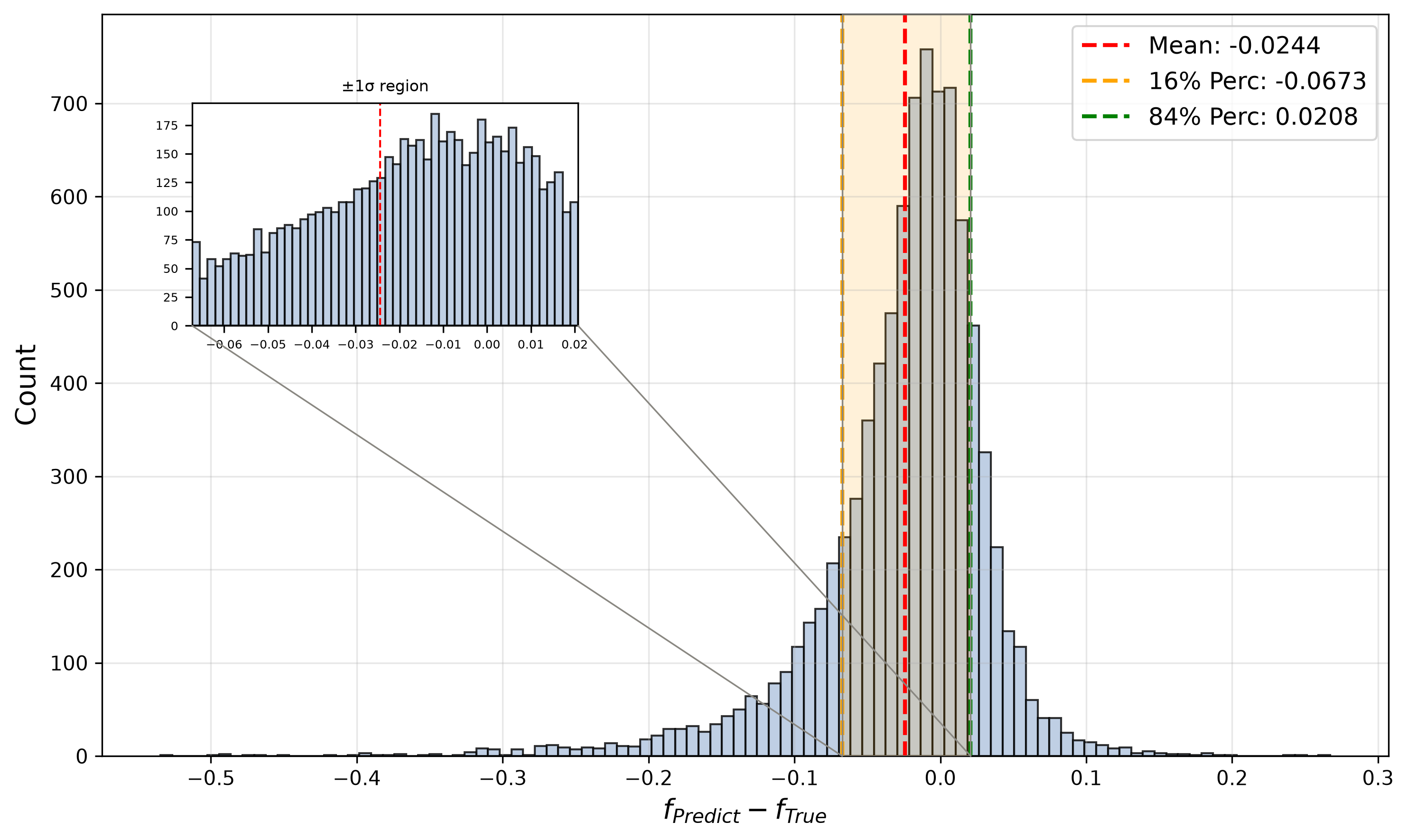}}
\subfigure{
\includegraphics[width=9cm,height = 5cm]{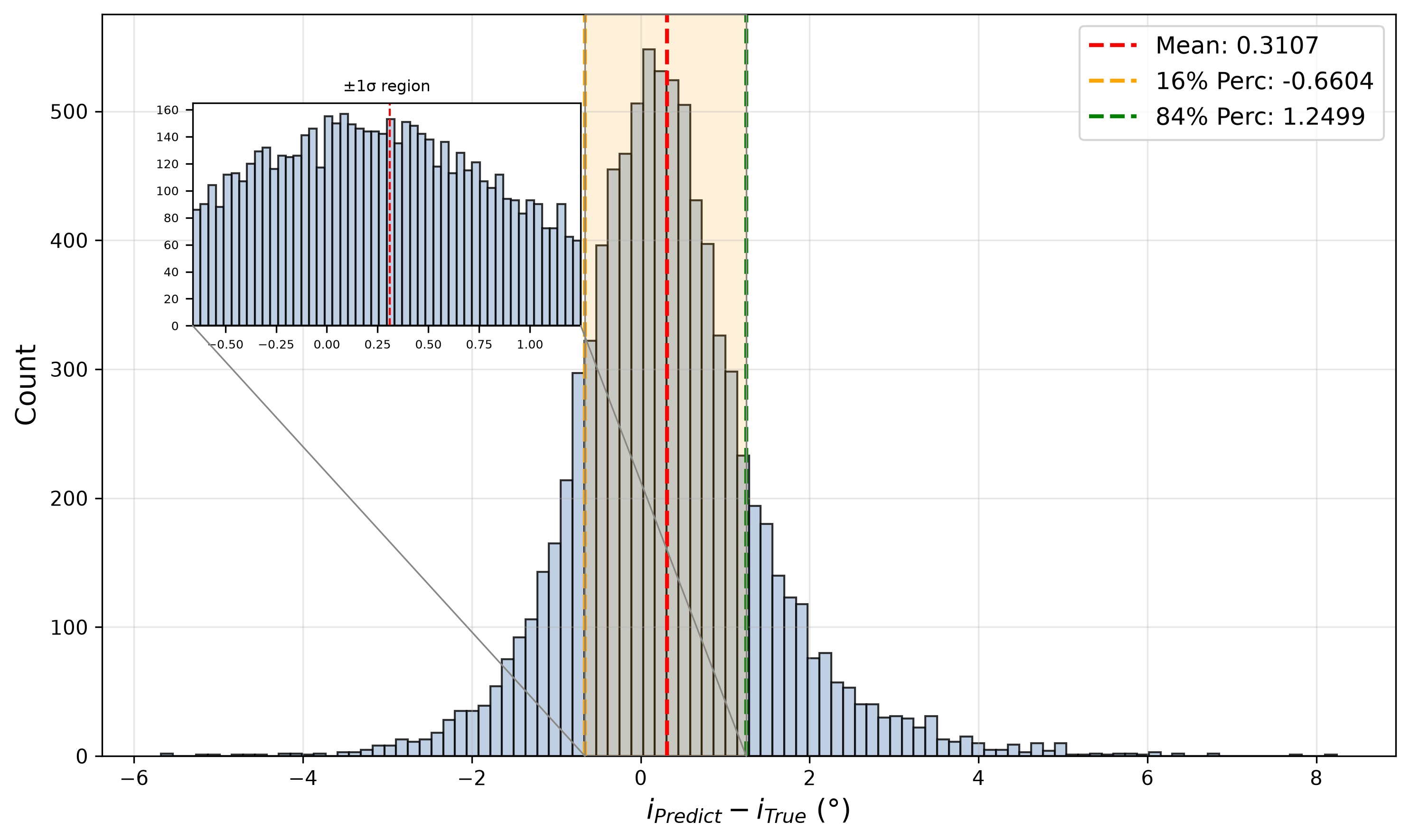}}
\subfigure{
\includegraphics[width=9cm,height = 5cm]{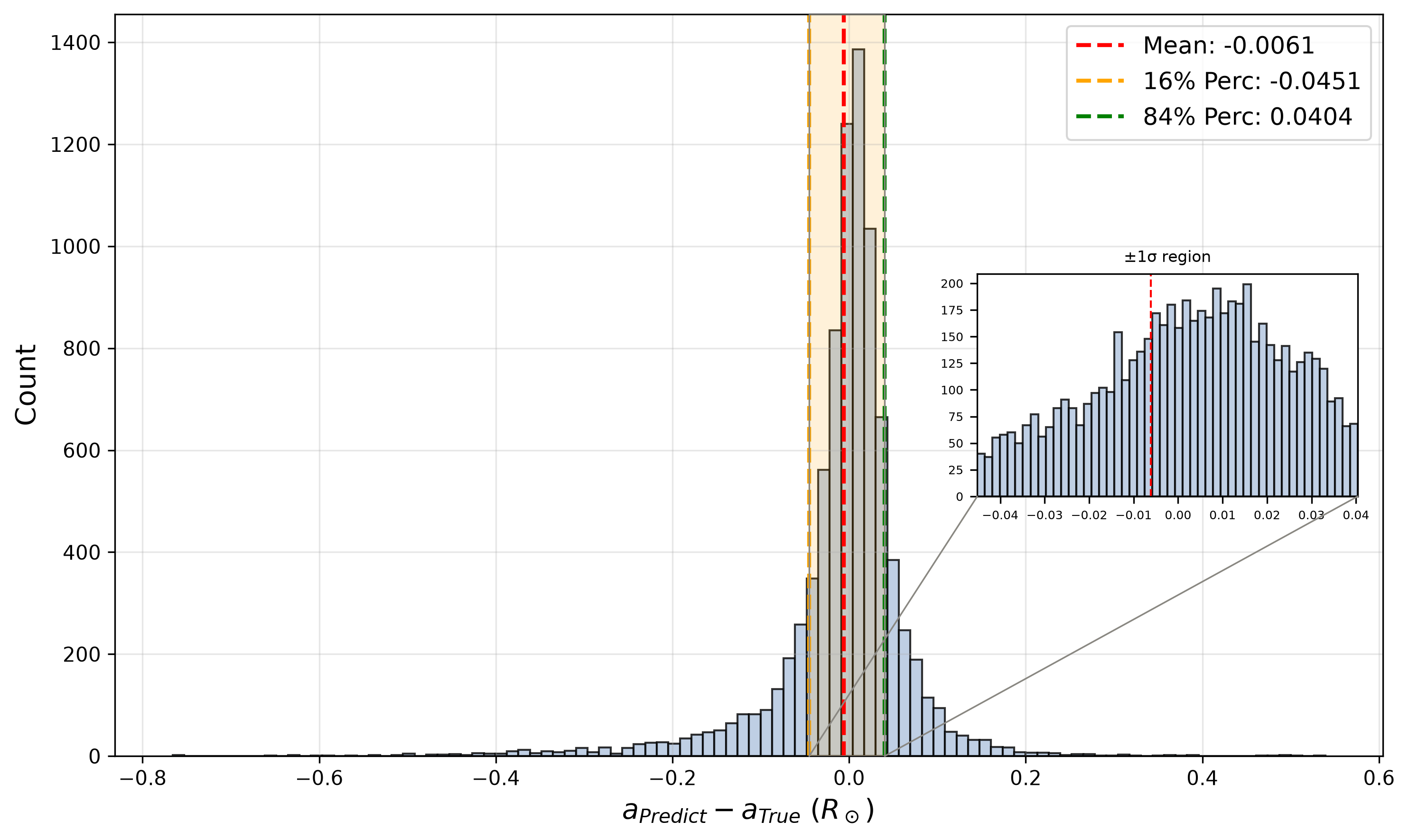}}
\subfigure{
\includegraphics[width=9cm,height = 5cm]{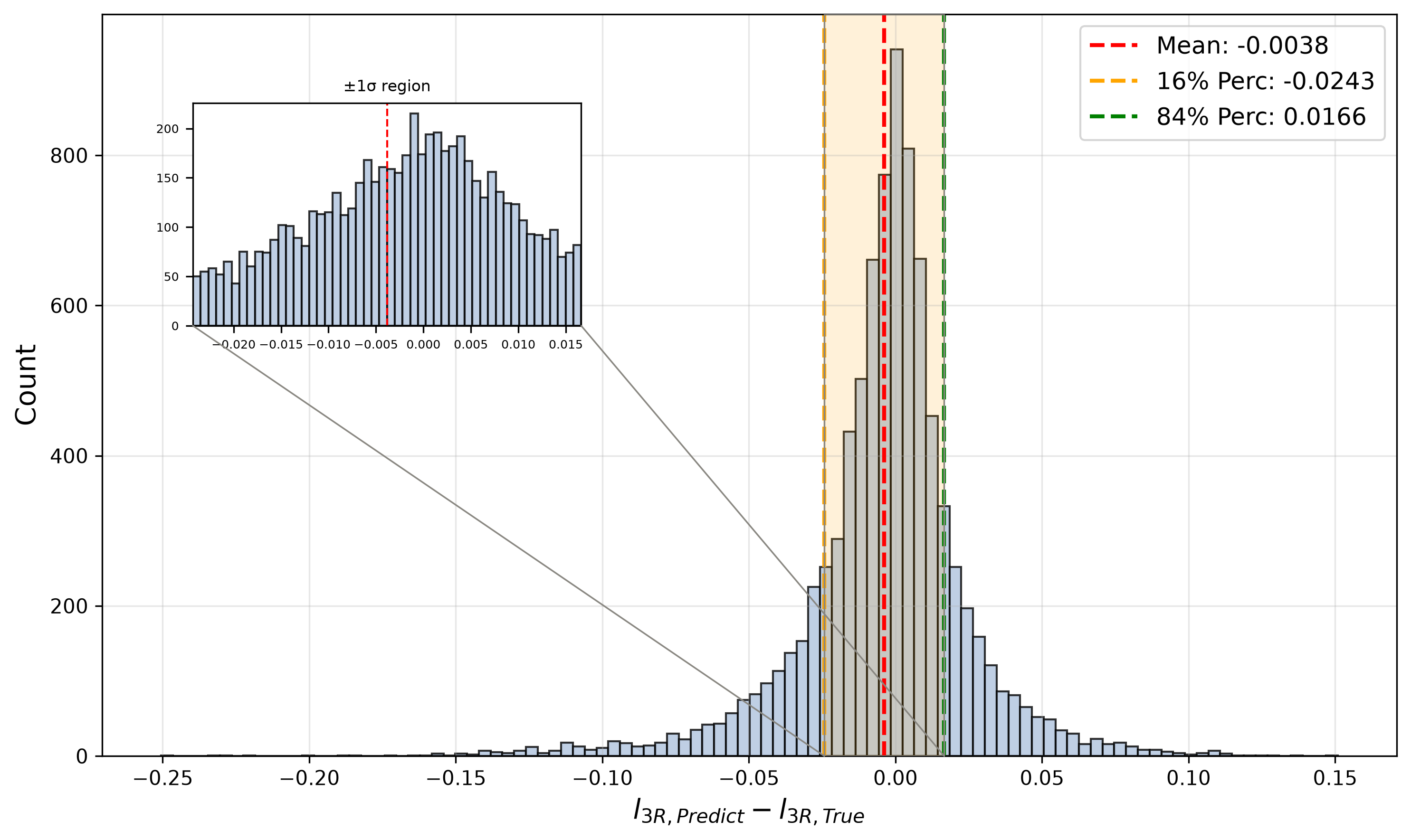}}
\subfigure{
\includegraphics[width=9cm,height = 5cm]{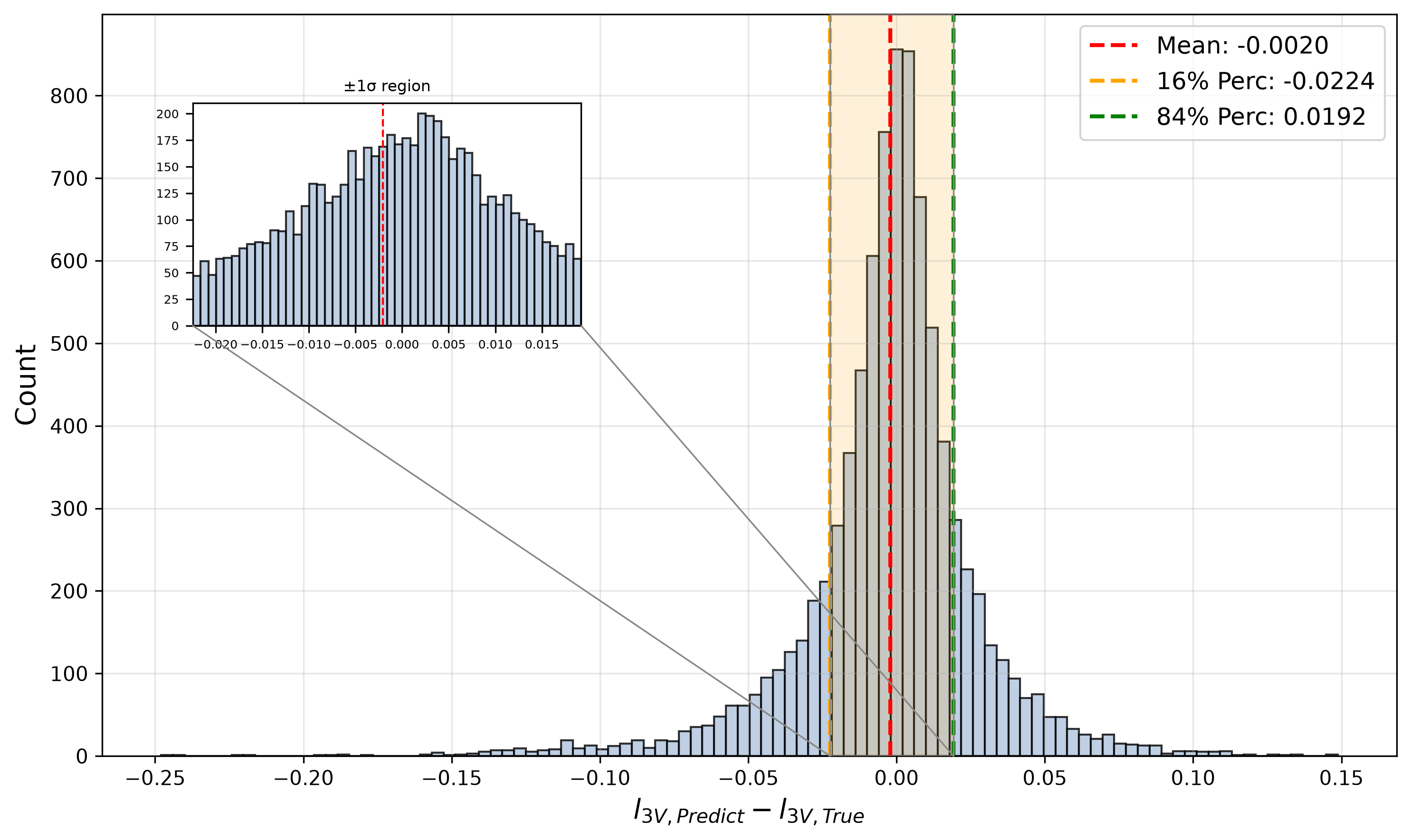}}
\subfigure{
\includegraphics[width=9cm,height = 5cm]{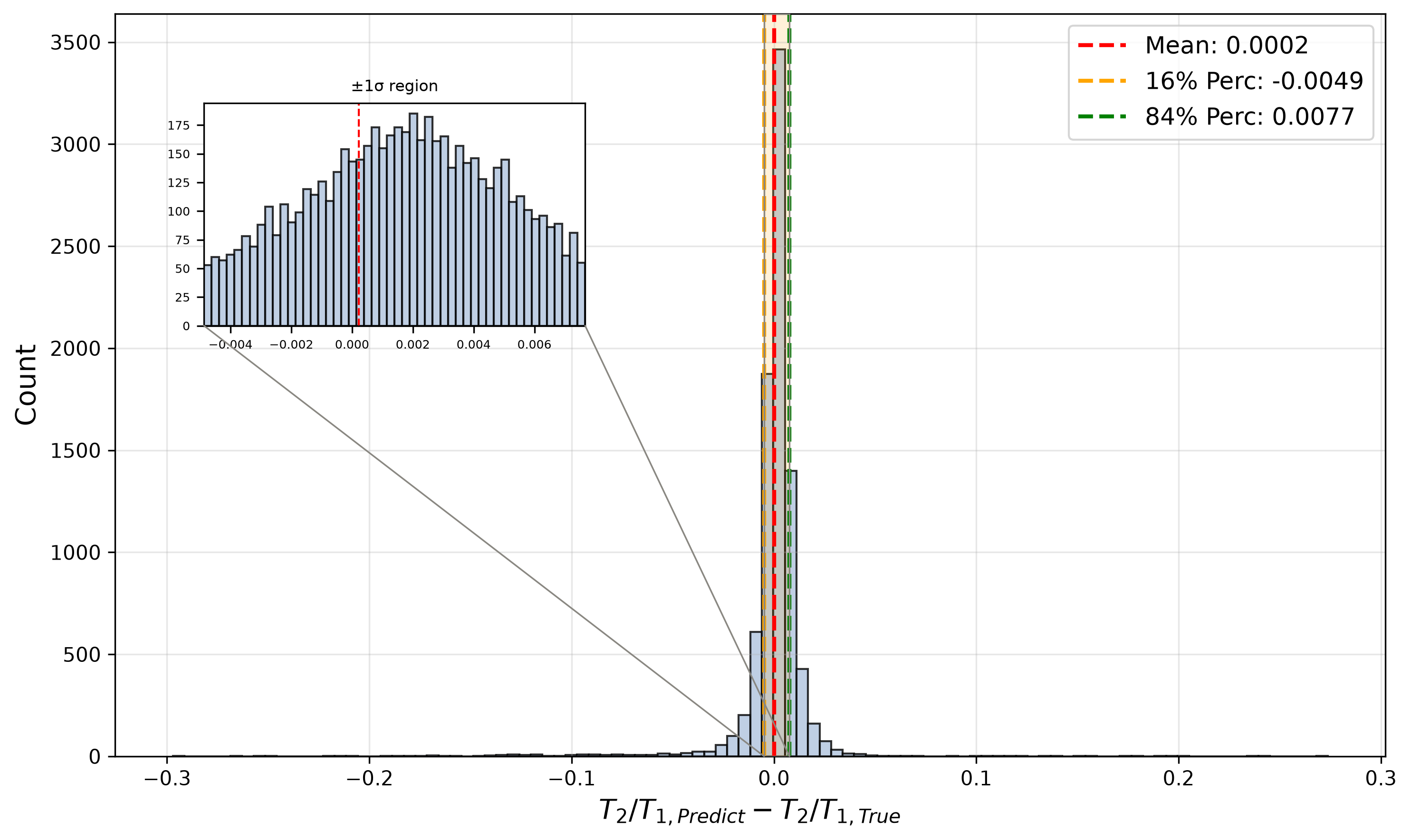}}
\subfigure{
\includegraphics[width=9cm,height = 5cm]{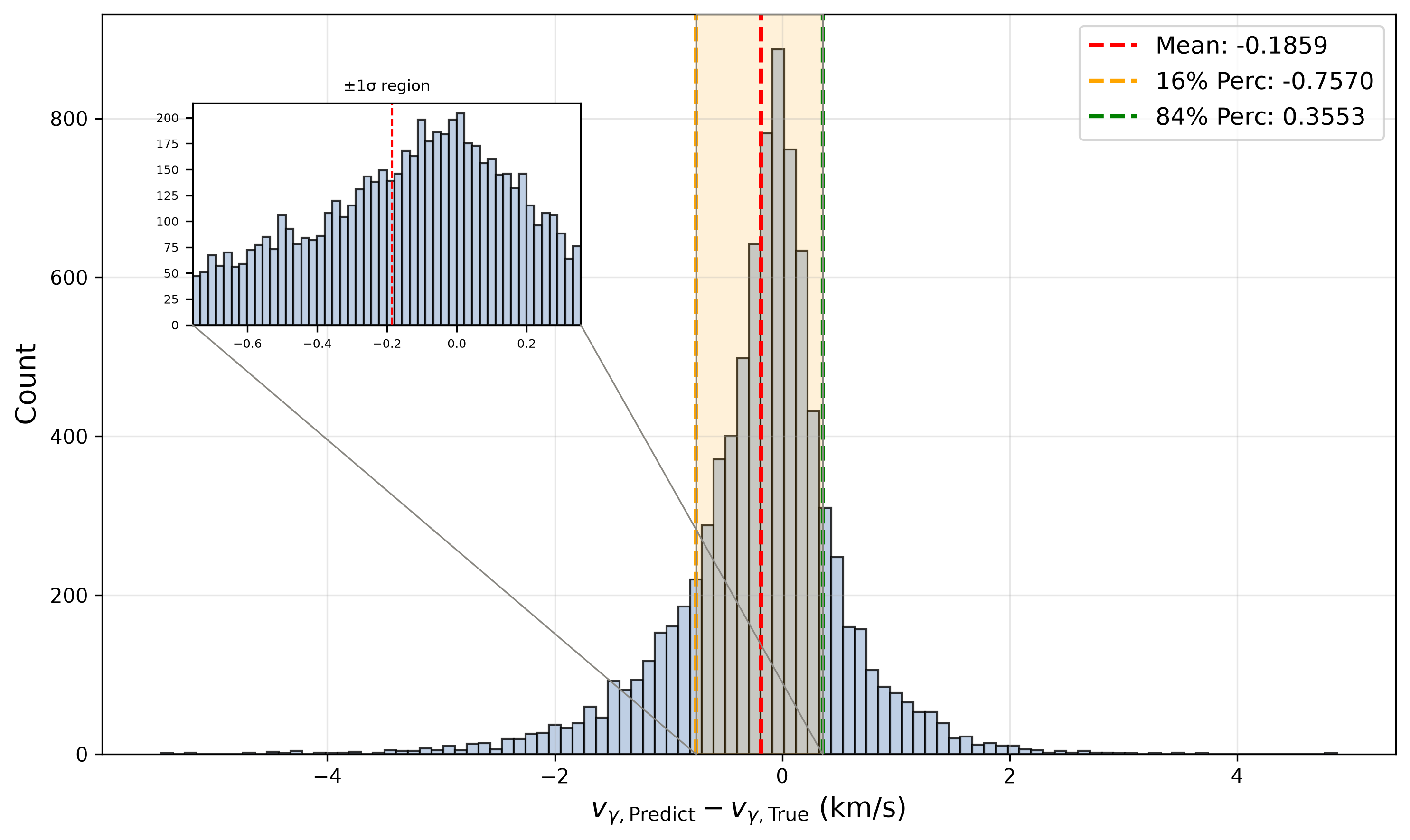}}
\caption{These figures illustrate the differences between the fitted parameters from the FCNN model and the true parameters. }
\label{fig:dis}
\end{figure*}

\begin{figure*}[htbp]

\subfigure{
\includegraphics[width=9cm,height = 5cm]{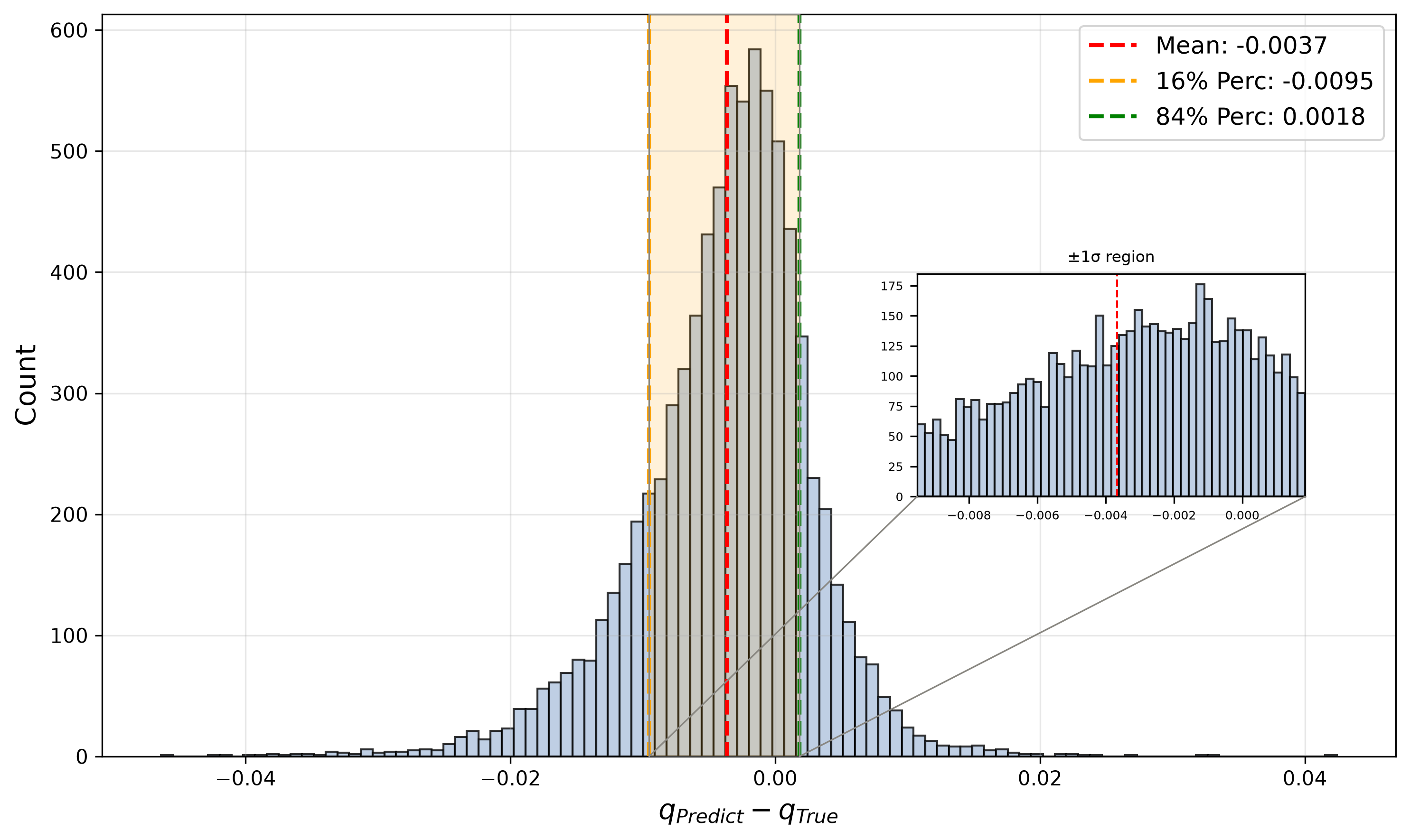}}
\subfigure{
\includegraphics[width=9cm,height = 5cm]{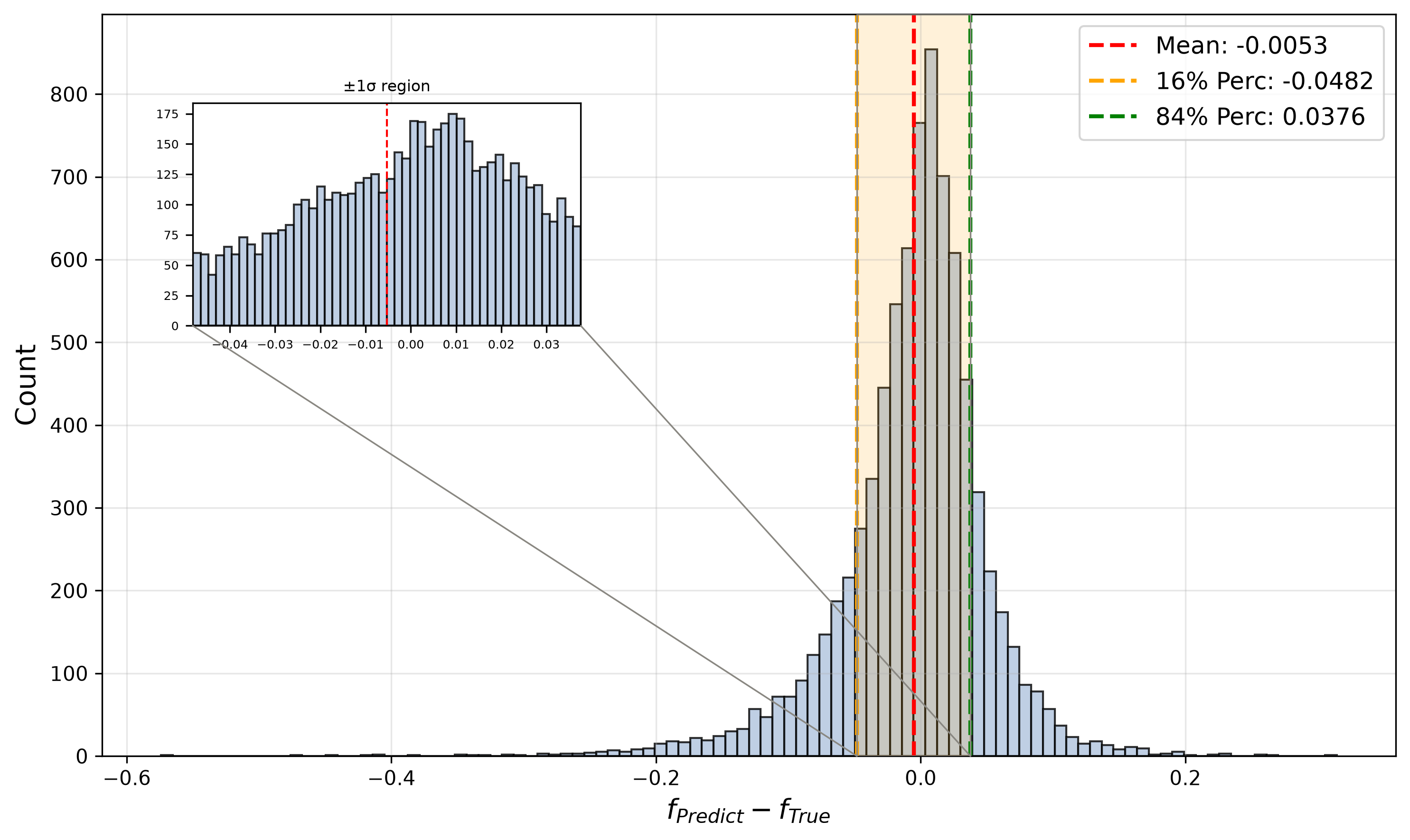}}
\subfigure{
\includegraphics[width=9cm,height = 5cm]{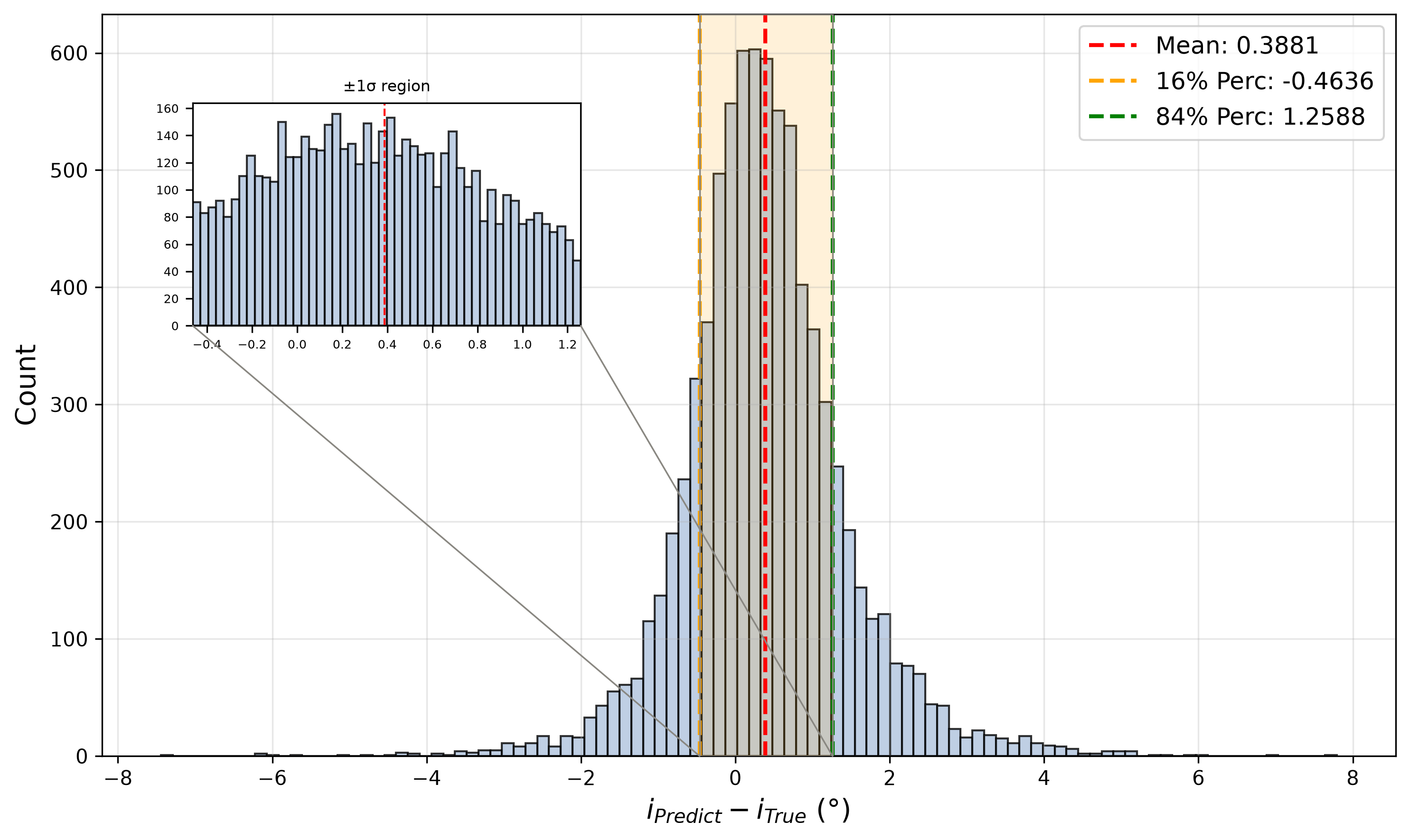}}
\subfigure{
\includegraphics[width=9cm,height = 5cm]{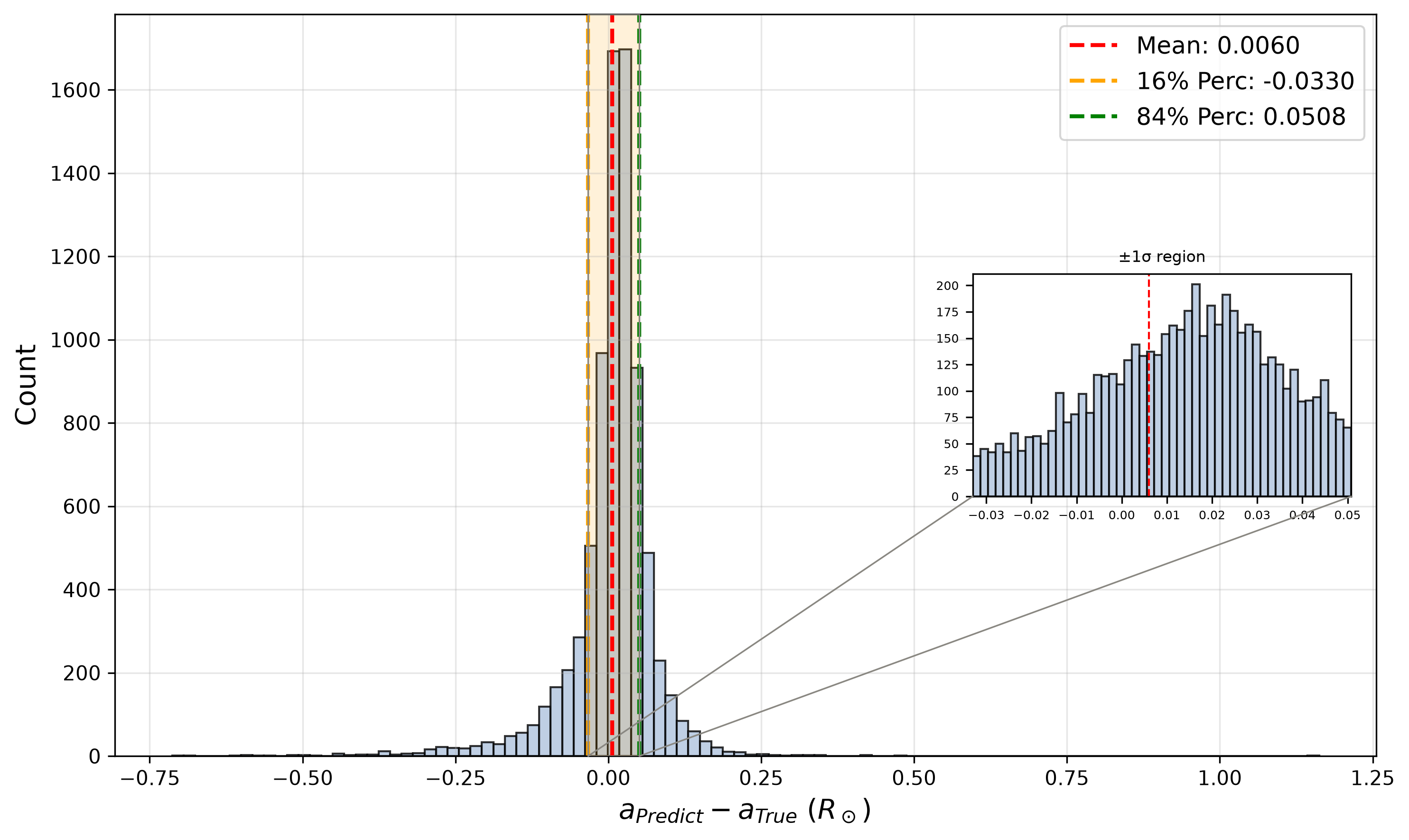}}
\subfigure{
\includegraphics[width=9cm,height = 5cm]{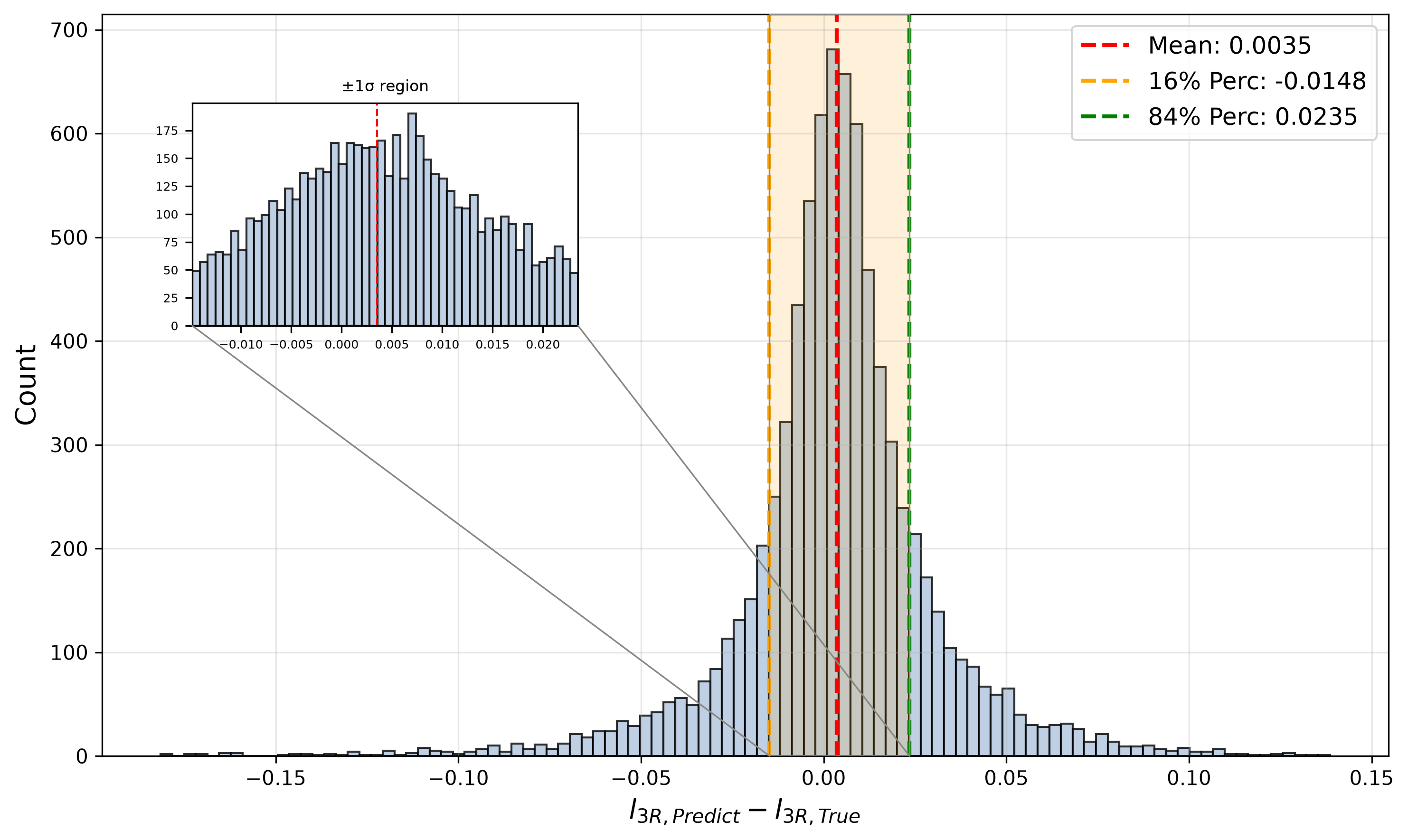}}
\subfigure{
\includegraphics[width=9cm,height = 5cm]{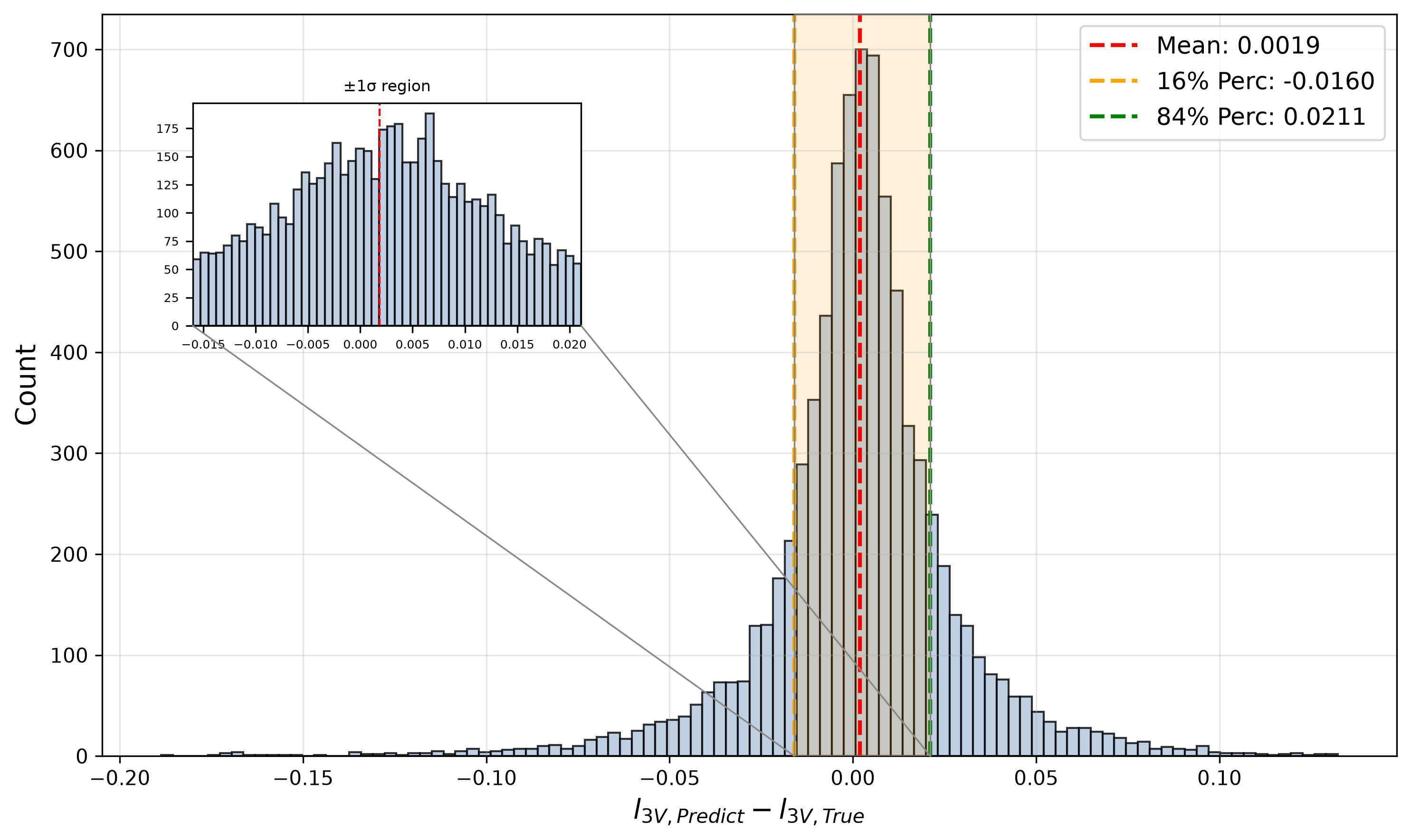}}
\subfigure{
\includegraphics[width=9cm,height = 5cm]{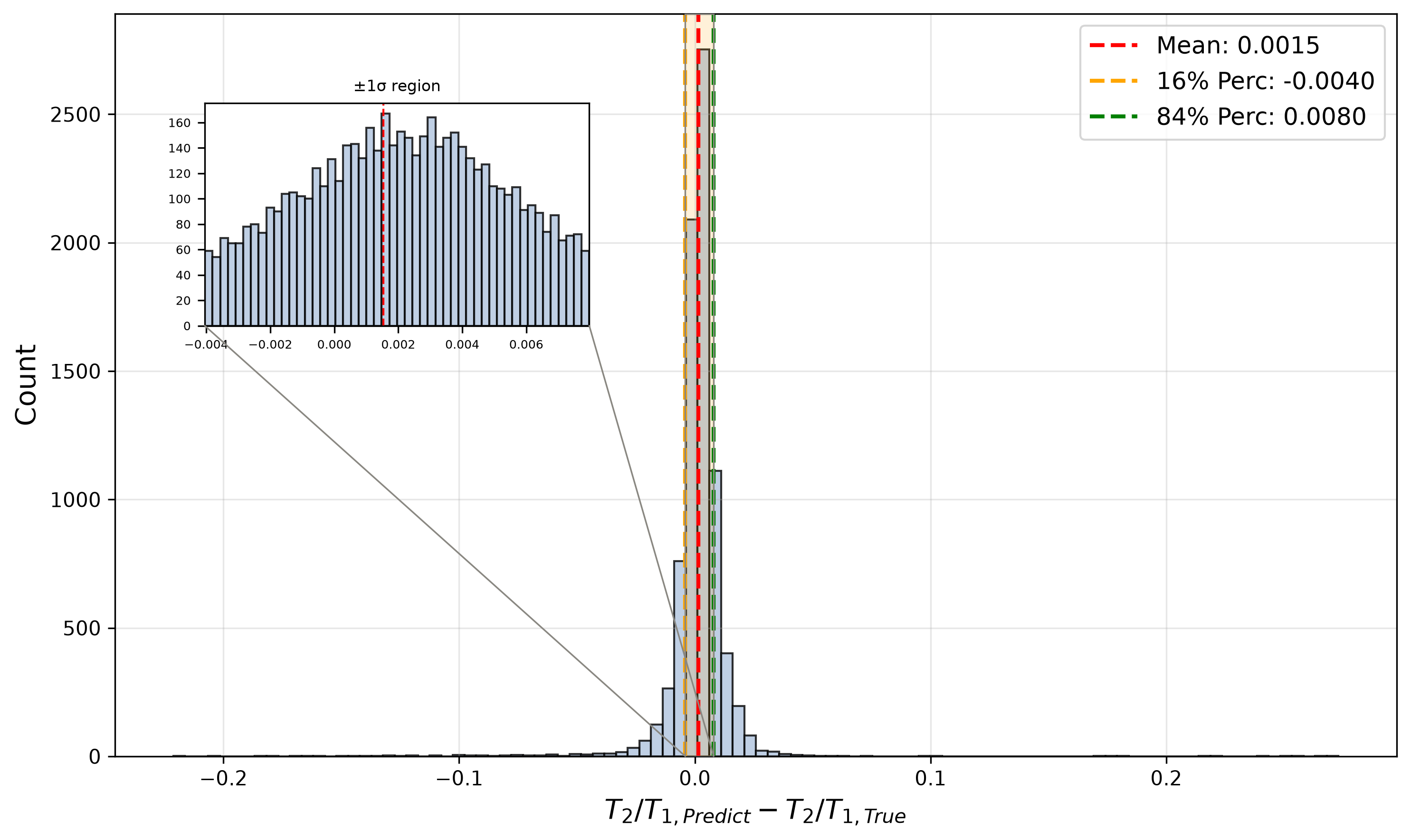}}
\subfigure{
\includegraphics[width=9cm,height = 5cm]{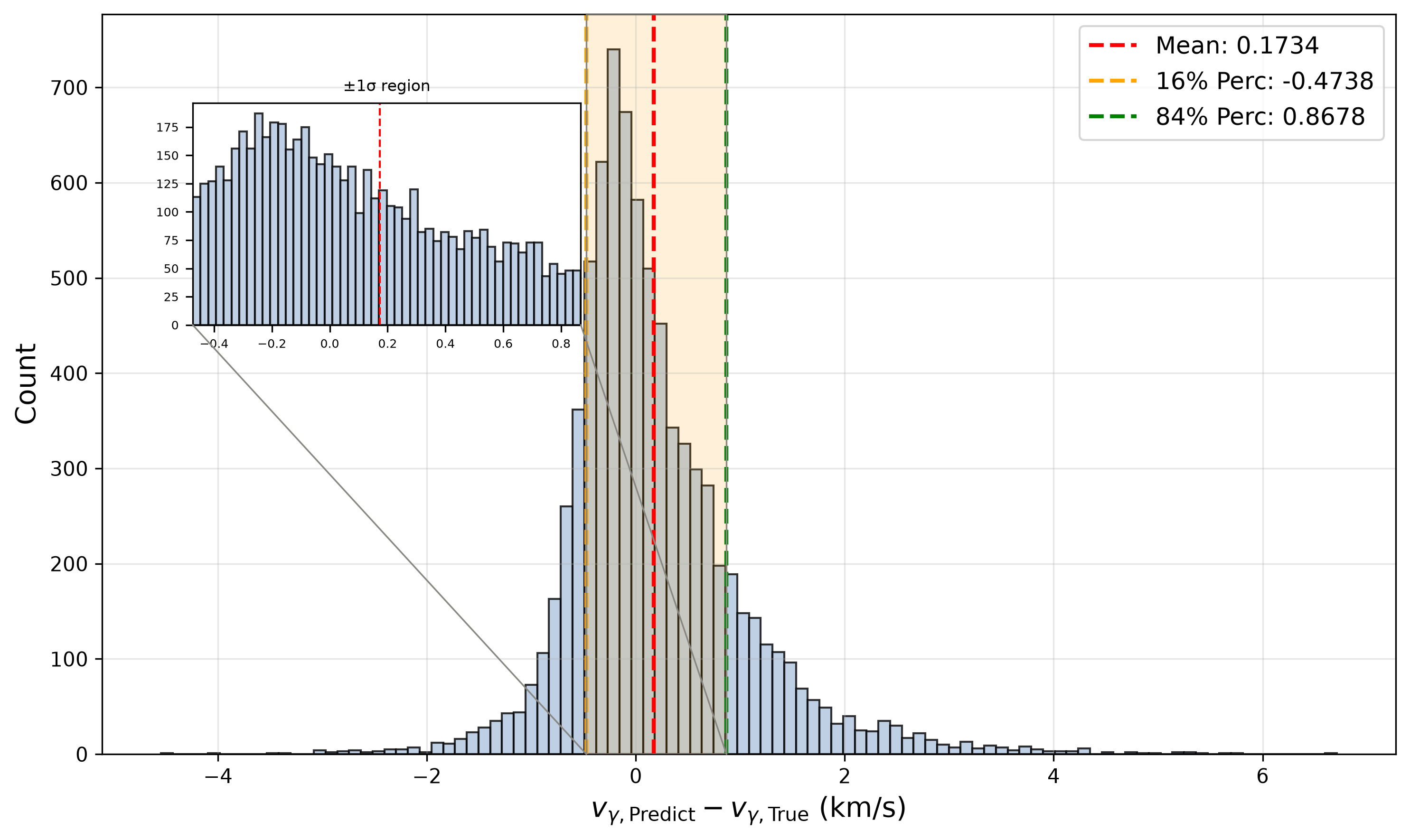}}
\caption{These figures illustrate the differences between the fitted parameters from the SIREN model and the true parameters. }
\label{fig:dis2}
\end{figure*}

We used two different models to perform the light-curve inversion independently. For each model, 10,000 synthetic datasets were used to evaluate the orbital solutions. We first assessed the overall fitting performance by requiring both the light-curve and radial-velocity fits to achieve $R^2>0.99$. Under this criterion, 9,974 systems (99.74\%) for the 1D model and 9,950 systems (99.50\%) for the SIREN model reached this level of fitting accuracy. 
For the subsequent parameter-recovery analysis, we adopted a more stringent criterion and retained only those systems for which both the light-curve and radial-velocity fits achieved $R^2>0.999$. This resulted in 8,676 systems for the 1D model and 8,073 systems for the SIREN model being retained, while 1,324 systems (13.24\%) and 1,927 systems (19.27\%), respectively, were not included in this high-precision subset. 
For the retained systems, orbital solution parameters were compared with their true values, and residual distributions for each parameter were analyzed (\Cref{fig:dis} and \Cref{fig:dis2}). Key statistical metrics are summarized in \Cref{tab:tabel1} and \Cref{tab:tabelsiren}. In \Cref{tab:tabel1} and \Cref{tab:tabelsiren}, the additional parameters include $l_{3R}$, which represents the third-light contribution in the R band, and $l_{3V}$ which represents the third-light contribution in the V band. For these two models, the difference in predictive performance is minimal, indicating that the SIREN model can achieve comparable accuracy to the FCNN with a substantially smaller number of parameters.

\renewcommand\arraystretch{1.2}
\begin{table*}[t]
\centering
\caption{\centering Residual distributions of the binary parameters derived with the FCNN model}
\label{tab:tabel1}
\begin{tabular}{lcccccccc}
\hline
Parameter 
& $\Delta q$ 
& $\Delta i$ ($^\circ$)& $\Delta(T_2/T_1)$ 
& $\Delta v_\gamma$ (km/s)
& $\Delta f$ 
& $\Delta a$ ($R_\odot$)
& $\Delta l_{3R}$ 
& $\Delta l_{3V}$ \\
\hline
max      & 0.1135 & 8.2449 & 0.2737 & 4.8776 & 0.2671 & 0.5393 & 0.1509 & 0.1488 \\
min      & -0.0367 & -5.6812 & -0.2971 & -5.4628 & -0.5345 & -0.7660 & -0.2506 & -0.2480 \\
mean     & -0.0004 & 0.3107 & 0.0002 & -0.1859 & -0.0244 & -0.0061 & -0.0038 & -0.0020 \\
$+1\sigma$ & 0.0044 & 1.2499 & 0.0077 & 0.3553 & 0.0208 & 0.0404 & 0.0166 & 0.0192 \\
$-1\sigma$ & -0.0053 & -0.6604 & -0.0049 & -0.7570 & -0.0673 & -0.0451 & -0.0243 & -0.0224 \\
\hline
\end{tabular}
\parbox{\textwidth}{\footnotesize
\hspace*{3.2cm}
Note. $\Delta X = X_{\rm Predict} - X_{\rm True}$.
}
\end{table*}

\renewcommand\arraystretch{1.2}
\begin{table*}[t]
\centering
\caption{\centering Residual distributions of the binary parameters derived with the SIREN model}
\label{tab:tabelsiren}
\begin{tabular}{lcccccccc}
\hline
Parameter 
& $\Delta q$ 
& $\Delta i$ ($^\circ$)& $\Delta(T_2/T_1)$ 
& $\Delta v_\gamma$ (km/s)
& $\Delta f$ 
& $\Delta a$ ($R_\odot$)
& $\Delta l_{3R}$ 
& $\Delta l_{3V}$ \\
\hline
max      & 0.0424 & 7.7946 & 0.2730 & 6.7119 & 0.3146 & 1.1624 & 0.1386 & 0.1323 \\
min      & -0.0464 & -7.4428 & -0.3062 & -4.5461 & -0.5742 & -0.7131 & -0.1815 & -0.1887 \\
mean     & -0.0037 & 0.3881 & 0.0015 & 0.1734 & -0.0053 & 0.0060 & 0.0035 & 0.0019 \\
$+1\sigma$ & 0.0018 & 1.2588 & 0.0080 & 0.8678 & 0.0376 & 0.0508 & 0.0235 & 0.0211 \\
$-1\sigma$ & -0.0095 & -0.4636 & -0.0041 & -0.4738 & -0.0482 & -0.0330 & -0.0148 & -0.0160 \\
\hline
\end{tabular}
\parbox{\textwidth}{\footnotesize
\hspace*{3.2cm}
Note. $\Delta X = X_{\rm Predict} - X_{\rm True}$.
}
\end{table*}

Overall, the orbital solution results show that the estimated values for most parameters are in close agreement with the true values. Nevertheless, a small subset of solutions exhibits significant deviations from the true parameters. Through further analysis, we found that such cases often involve one or more parameters taking extreme values. Moreover, these parameters tend to exert relatively limited influence on the shape of the light curve (for instance, the fill-out factor may display considerable discrepancy despite minimal impact on the overall fit). These findings indicate that the proposed program and solution strategy are sufficiently robust for large-scale batch analysis of contact binary datasets. In exceptional cases, accurate results may require customized approaches, such as first constraining initial parameter ranges via alternative methods and then applying localized MCMC optimization within tighter bounds.

\begin{figure}
    \centering
    \includegraphics[width=1\linewidth]{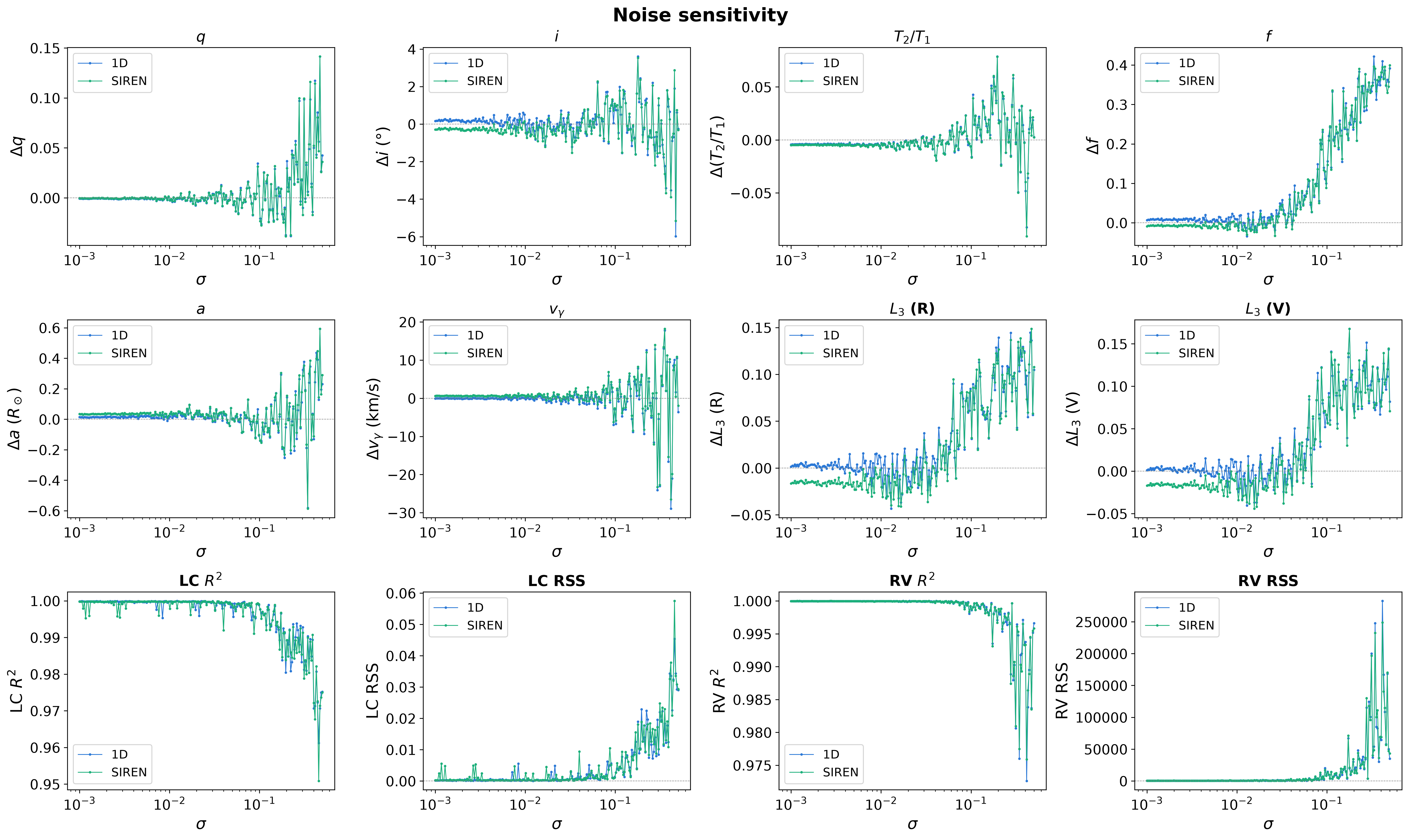}
    \caption{The figure illustrates the variations in the parameter deviations as a function of the noise level, together with the corresponding $R^2$ and RSS values obtained from comparisons between the reconstructed and original theoretical curves.}
    \label{fig:add_noise}
\end{figure}

\begin{figure}
    \centering
    \includegraphics[width=1\linewidth]{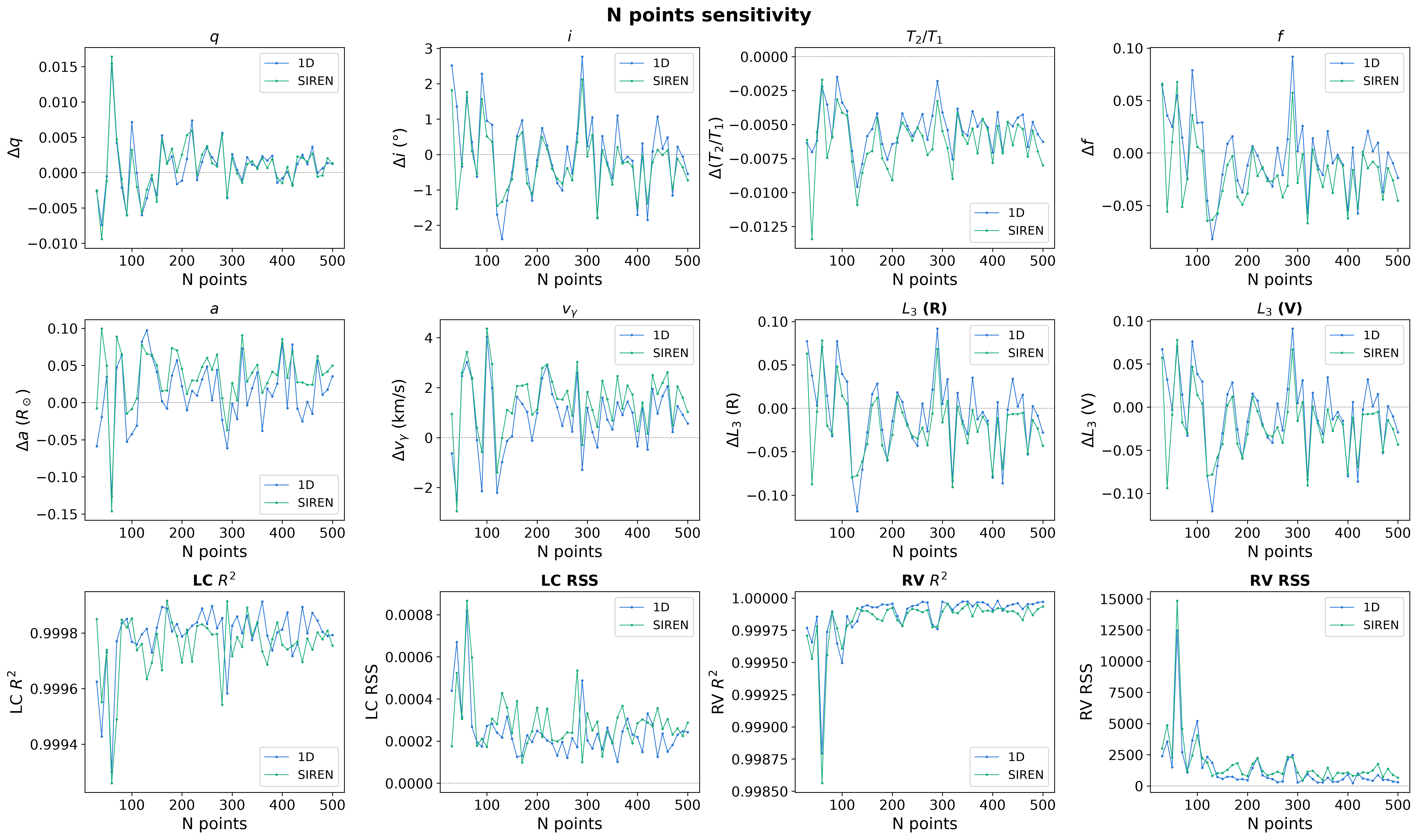}
    \caption{The figure illustrates the variations in the parameter deviations with different numbers of data points, together with the corresponding $R^2$ and RSS values obtained by comparing the final reconstructed curves with the original theoretical curves. }
    \label{fig:add_point}
\end{figure}

To evaluate the applicability of the proposed framework to real observational data, a series of simulated test datasets were constructed from two perspectives based on theoretical contact binary systems with known parameters, together with their corresponding synthetic light curves and radial velocity curves.
First, to quantitatively assess the influence of observational noise on orbital solution accuracy, 200 values of the noise standard deviation $\sigma$ were selected with logarithmically uniform spacing between 0.001 and 0.5. Zero-mean Gaussian random noise was then injected into the synthetic curves. Given the substantial differences in numerical scale and dynamic range between light curves and radial velocity curves, the noise standard deviation was normalized by the amplitude of each curve type separately. This normalization ensured that the same noise level produced comparable relative perturbations in both datasets.
Second, to investigate the effect of sampling density on solution accuracy, the synthetic curves were resampled using sample sizes ranging from 30 to 500 points with an interval of 10 points. The sampling locations were randomly distributed across the orbital phase rather than uniformly spaced, thereby mimicking the non-uniform phase coverage commonly encountered in real observations.
The proposed framework was subsequently applied to each degraded dataset. The deviations between the derived parameters and their corresponding true values were adopted as the primary accuracy metric, enabling a systematic assessment of the impacts of noise level and sampling density on the orbital solution results.
In addition to the parameter estimation errors, $R^2$ and RSS between the reconstructed and original theoretical curves were computed as complementary indicators of solution quality. The variations of these metrics under different noise levels and sampling densities, as well as their effects on the computational accuracy of individual parameters, are displayed  in  \Cref{fig:add_noise} and \Cref{fig:add_point}.

The results obtained by introducing different levels of noise demonstrate that the proposed framework can accurately determine the true parameters when the noise level is below $\sigma$=0.1, indicating its strong applicability to real observational data affected by noise. Meanwhile, the results obtained by varying the number of data points show that the computed parameters remain in good agreement with the true values even when only a limited number of data points are available. Taken together, these results demonstrate that the proposed framework exhibits good robustness to both observational noise and sparse sampling, supporting its applicability to real observational scenarios. 

The program has been publicly released and is accessible on GitHub\footnote{ \url{https://github.com/0batsy0/CBW}} and the
Zenodo data repository (\dataset[https://doi.org/10.5281/zenodo.21994046]{CBW v1.0.1}).

\section{CONCLUSIONS} \label{sec:CONCLUSIONS}

In this study, we trained two FCNN models with 1D and 2D output representations, respectively, and a SIREN model to rapidly generate theoretical radial velocity curves.  A total of 50,000 parameter sets were randomly sampled within specified ranges, and the corresponding radial velocity curves were synthesized using PHOEBE. These synthetic curves were used as the training dataset, enabling the models to efficiently generate theoretical radial velocity curves based on user-input parameters.

To evaluate the accuracy of the models, 10,000 parameter sets were randomly generated, and the corresponding radial velocity curves were generated using both PHOEBE and the trained neural networks. The results indicate that all three models exhibit comparable performance. Considering the simplicity of the model architecture, the 1D architecture was adopted instead of the 2D architecture. Meanwhile, the FCNN is approximately 11 times larger than the SIREN model, demonstrating that the SIREN architecture can achieve comparable performance with a much more compact structure. However, due to the use of sinusoidal activation functions, which involve relatively complex computations, the SIREN model tends to require greater computational resources during inference. 

Subsequently, we integrated the trained models into the framework proposed by \cite{TCW2025}, providing a unified tool for the joint orbital solution of radial velocity and photometric light curves. We generated 10,000 pairs of light curves and radial velocity curves using PHOEBE and analyzed them with the proposed models. The results show that the derived orbital solutions achieve very high accuracy. Analysis of a small number of poorly fitted cases indicates that these typically involve extreme parameter values or parameters with limited influence on the light curve morphology.

To further assess the dependence of the proposed framework on data quality and the sensitivity of the recovered parameters, a series of experiments were conducted by introducing different noise levels and varying the number of data points. The results indicate that the framework remains robust under both noisy and sparsely sampled conditions, demonstrating its reliability and suitability for practical applications involving real observational data. 

Overall, these results demonstrate that the proposed models and solution strategy are robust for the vast majority of targets. For a few special cases, customized solution strategies may be required, such as refining the initial parameter ranges before performing local MCMC optimization. Both models have been made publicly available on GitHub and Zenodo to facilitate further use by the research community.

 \section*{ACKNOWLEDGEMENTS}
 We are grateful to the anonymous reviewer for the careful evaluation of our manuscript and for providing valuable insights that have helped us further strengthen our work. 
This work is supported by the National Natural Science Foundation of China (NSFC; No. 12273018), the Taishan Scholars Young Expert Program of Shandong Province, the Qilu Young Researcher Project of Shandong University, the Young Data Scientist Project of the National Astronomical Data Center. The calculations in this work were carried out at the Supercomputing Center of Shandong University, Weihai.

\bibliography{sample631}{}
\bibliographystyle{aasjournal}



\end{document}